\documentclass[aps,prl,superscriptaddress,reprint,showpacs,floatfix]{revtex4-2}
\usepackage{graphicx}
\usepackage{dcolumn}
\usepackage{bm}
\usepackage{xcolor}
\usepackage{relsize}
\usepackage{amsmath}
\usepackage{amsfonts}
\usepackage{amssymb}
\usepackage{mathtools}
\usepackage{braket}
\usepackage{gensymb}
\usepackage{url}
\usepackage{mathrsfs}
\usepackage{enumitem}
\usepackage{physics}
\setlist[itemize]{label={}}

\usepackage[colorlinks = true,
            linkcolor = blue,
            urlcolor  = blue,
            citecolor = blue,
            anchorcolor = blue]{hyperref}
\usepackage{footmisc}
\usepackage[utf8]{inputenc}

\begin{document}

\preprint{APS/123-QED}

\title{Ground-State Selection by Pure Energy Relaxation in Polariton Condensates}

\author{D. A. Saltykova}
\affiliation{ITMO University, St. Petersburg 197101, Russia}

\author{A. V. Yulin}
\affiliation{ITMO University, St. Petersburg 197101, Russia}

\author{I. A. Shelykh}

\affiliation{Science Institute, University of Iceland, Dunhagi 3, IS-107 Reykjavik, Iceland}

\begin{abstract}
We study nonequilibrium mode selection in dissipative exciton-polariton condensates incoherently pumped through an excitonic reservoir in the presence of pure energy relaxation. For a confined system in which a vortex mode is selected at threshold, we show that energy relaxation qualitatively changes the condensation scenario: as the pump increases, the asymptotic state evolves from a vortex condensate to a rotating mixed state and then to a ground-state condensate. Pure energy relaxation thus destabilizes condensation into excited states and promotes ground-state selection.

\end{abstract}

\maketitle

\textit{Introduction.}
Cavity polaritons are hybrid light-matter quasiparticles arising in semiconductor microcavities operating in the strong-coupling regime \cite{Kavokin2017_OxfPr}. They exhibit a range of remarkable collective quantum phenomena at surprisingly high temperatures \cite{Carusotto2013}. The most prominent are polariton superfluidity \cite{Amo2009,Lerario2017} and polariton lasing \cite{Christopoulos2007,Schneider2013}, which are related to the formation of macroscopically populated coherent polariton states, known as polariton Bose--Einstein condensates (BECs) \cite{Kasprzak2006,Balili2007}. The possibility of achieving high-temperature polariton condensation is defined by a combination of factors, including an extremely small mass of cavity polaritons (about $10^{-5}$ of the free-electron mass), their macroscopically long coherence length (on the millimeter scale) \cite{Ballarini2017}, a high degree of optical nonlinearity stemming from efficient exciton-exciton interactions \cite{Glazov2009,Estrecho2019,Snoke2023}, and the possibility of fast thermalization in the system \cite{Malpuech2002,Richard2005,Sun2017}.

Polariton BECs, unlike atomic ones, have an inherently dynamic nature. Although the formation of a quasi-thermal polariton distribution has been routinely observed \cite{Richard2005,Kasprzak2007,Sun2017}, the condensation process is not driven by thermalization alone, but by its delicate interplay with external pumping and decay processes, and is strongly influenced by the coupling of the condensate to a reservoir of incoherent excitons \cite{Galbiati2012,Estrecho2018}, so that condensation into excited states can be observed \cite{Galbiati2012,Sedov2021,Aladinskaia2023}. A coherent phenomenological description of this phenomenon must therefore account for the balance between the different dissipation mechanisms in the system.

The effects of incoherent pumping and the coupling between polaritons and incoherent excitons are usually described within the framework of the Wouters--Carusotto model, which introduces the coupling between the macroscopic wavefunction of the condensate and the incoherent excitonic reservoir, and was first formulated for scalar polaritons \cite{WoutersCarusotto2007} and then generalized to the spinor case \cite{Borgh2010}. However, this model completely neglects energy relaxation within the condensate, and within its framework condensation always occurs into the state with the optimal balance between its quality factor and the scattering rate, independently of its energy. In the present paper, we revise the existing theory of dynamic polariton condensation by introducing a term describing pure energy relaxation, following our recent publication \cite{Saltykova2025}. We apply our theory to the study of mode competition in a spatially confined system, focusing on the interplay between the ground state and excited vortex-like states. We demonstrate that the presence of energy relaxation dramatically changes the dynamics, leading to the appearance of novel types of rotating states and driving condensation into the ground state under strong pumping.

\textit{Theoretical Model and Perturbative Mode-Competition Analysis.}
In the experimentally relevant regime of fast reservoir relaxation, the excitonic reservoir can be adiabatically eliminated \cite{Keeling2008}, reducing the coupled condensate-reservoir dynamics to an effective generalized Gross-Pitaevskii equation for the polariton order parameter $\psi(\mathbf r,t)$:
\begin{widetext}
\begin{equation}
i\frac{\partial \psi}{\partial t}
=\left[
-\frac{\hbar \nabla^{2}}{2m_{LP}}
+ V_{\mathrm{eff}}(|\psi|^{2})
+ i\,\Gamma_{\mathrm{eff}}(|\psi|^{2})
\right]\psi
+\frac{i}{2}\lambda \psi \left(\psi^{*}\nabla^{2}\psi-\psi\nabla^{2}\psi^{*}\right),
\label{GPE}
\end{equation}
\end{widetext}
where $\psi(\mathbf r,t)$ is the condensate wavefunction, $m_{LP}$ is the effective polariton mass, and $\lambda$ is the energy-relaxation coefficient \cite{Saltykova2025}. Here and below, all energies are expressed in frequency units, i.e., divided by $\hbar$. The effective conservative and dissipative nonlinearities are given by
\begin{widetext}
\begin{equation}
V_{\mathrm{eff}}(|\psi|^{2}) = V_{\mathrm{trap}}(r)+g|\psi|^{2} + \frac{2\tilde g}{\gamma_R } \left(1- \frac{R}{\gamma_R}|\psi |^2 \right) \eta(r)\mathcal{P},
\qquad
\Gamma_{\mathrm{eff}}(|\psi|^{2}) = \frac{1}{2}\left(\frac{R}{\gamma_R} \left(1- \frac{R}{\gamma_R}|\psi |^2\right) \eta(r)\mathcal{P} - \gamma\right).
\label{eq:VeffGammaeff}
\end{equation}
\end{widetext}

The reduced description is expected to be quantitatively reliable as long as the reservoir relaxation remains the fastest process and the condensate density stays within the weak-saturation regime. In particular, when the pump-induced blueshift becomes sufficiently strong to substantially modify the underlying linear modes, the perturbative analysis developed below should be regarded as qualitative rather than quantitative.
Here, $g$ and $\tilde g$ describe polariton-polariton and polariton-reservoir interactions, respectively. The term $V_{\mathrm{trap}}(r)$ is the external trapping potential, which may originate, for example, from microstructuring. The pump is written as $P(\mathbf r)=\mathcal P\,\eta(r)$, where $\mathcal P$ is the pump amplitude and $\eta(r)$ is the normalized spatial pump profile, $\max_r|\eta(r)|=1$. The parameter $\gamma$ is the intrinsic condensate decay rate, while $\gamma_R$ is the reservoir decay rate, and $R$ characterizes stimulated scattering from the reservoir into the condensate.

The terms inside the square brackets in Eq.~\eqref{GPE} describe the standard driven-dissipative polariton dynamics, including the trap, interactions, and reservoir-induced gain and saturation. The last term, proportional to $\lambda$, represents pure energy relaxation within the condensate. It originates from polariton-phonon scattering, conserves particle number, and favors lower-energy modes \cite{Saltykova2025}. The derivation of Eq.~\eqref{GPE} from the full coupled condensate-reservoir model is given in the Supplementary Material, Note 1.

It is well known that condensation can be viewed as a competition between growing modes; see, for example, Refs.~\cite{benkert1991controlled,yulin2016spontaneous,yulin2023vorticity}. This competition arises because each growing mode depletes the gain available both to itself and to the other modes. If the depletion of the other modes is stronger than the self-depletion, the dominant mode suppresses its competitors. The surviving modes determine the properties of the stationary condensate, in particular its topological charge.

To illustrate the importance of energy relaxation, we first employ a perturbative approach that describes the dynamics in terms of the amplitudes of interacting modes. We consider an axially symmetric conservative potential that is sufficiently deep to treat dissipative and nonlinear effects as small perturbations. Accordingly, this description is valid provided that the pump-induced blueshift and gain saturation do not substantially reshape the linear eigenmodes of the trap.  We focus on the interaction between the fundamental mode and the lowest-lying vortex mode with azimuthal index $m=1$. The mode with $m=-1$ may also be included, but the analysis shows that the two vortex modes with $m=\pm1$ compete strongly with each other, so that only one of them can survive; see the Supplementary Material, Note 2 for details. Higher-order modes experience significant losses, which can be ensured by an appropriate choice of the pump profile, and can therefore be neglected.

To analyze mode competition, we expand the condensate field in two orthonormal eigenmodes of the corresponding conservative problem,
\begin{align}
\psi(\mathbf r,t)
&=A_G(t)\,\phi_G(\mathbf r)\,e^{-i\Omega_G t}
+ A_v(t)\,\phi_v(\mathbf r)\,e^{-i\Omega_v t},
\label{eq:wavefunct}
\end{align}
with $\phi_G(\mathbf r)=y_G(r)$ and $\phi_v(\mathbf r)=y_v(r)e^{i\theta}$.
By orthonormality,
\begin{equation}
A_j(t)e^{-i\Omega_j t}
=
\int_{\mathbb R^2}\phi_j^*(\mathbf r)\,\psi(\mathbf r,t)\,d^2r,
\qquad j\in\{G,v\}.
\end{equation}
Using the expansion \eqref{eq:wavefunct}, we project Eq.~\eqref{GPE} onto $\phi_G$ and $\phi_v$. Keeping only resonant terms and treating dissipation and nonlinearities perturbatively, we obtain coupled equations for the mode amplitudes. The corresponding intensities,
\[
I_G=|A_G|^2,\qquad I_v=|A_v|^2,
\]
then satisfy
\begin{align}
\dot I_G &= 2I_G\Bigl(\Gamma_G-\sigma_{GG}\mathcal P\, I_G-\bigl(2\sigma_{Gv}\mathcal P-\rho_{vG}\bigr)I_v\Bigr),
\label{eq:IG_dyn}\\
\dot I_v &= 2I_v\Bigl(\Gamma_v-\sigma_{vv}\mathcal P\, I_v-\bigl(2\sigma_{Gv}\mathcal P+\rho_{vG}\bigr)I_G\Bigr).
\label{eq:Iv_dyn}
\end{align}
Here
\begin{equation}
\Gamma_j=-\tilde\Gamma_j+\epsilon_j\mathcal P,\qquad j\in\{G,v\},
\end{equation}
are the effective linear growth rates of the modes, with
\begin{equation}
\tilde\Gamma_j=\frac{1}{2}\int_{\mathbb R^2}\gamma\,|\phi_j(\mathbf r)|^2\,d^2r,
\end{equation}
and
\begin{equation}
\epsilon_j=\frac{R}{2\gamma_R}\int_{\mathbb R^2}\eta(r)\,|\phi_j(\mathbf r)|^2\,d^2r.
\end{equation}
The nonlinear gain-saturation coefficients are
\begin{equation}
\sigma_{jj}=\frac{R^2}{2\gamma_R^2}\int_{\mathbb R^2}\eta(r)\,|\phi_j(\mathbf r)|^4\,d^2r,
\qquad j\in\{G,v\},
\end{equation}
and
\begin{equation}
\sigma_{Gv}=\frac{R^2}{2\gamma_R^2}\int_{\mathbb R^2}\eta(r)\,|\phi_G(\mathbf r)|^2|\phi_v(\mathbf r)|^2\,d^2r,
\end{equation}
the latter describing cross-saturation between the modes. The effect of pure energy relaxation is taken into account by
\begin{equation}
\rho_{vG}=
\lambda\,\frac{m_{LP}}{\hbar}\,(\Omega_v-\Omega_G)\,
\int_{\mathbb R^2}|\phi_G(\mathbf r)|^2\,|\phi_v(\mathbf r)|^2\,d^2r.
\end{equation}

The theory confirms the intuitive result that the growth rate is determined by the overlap of the eigenmode with the pump profile and the losses. This makes it possible to selectively excite the state corresponding to a desired eigenmode. Another important observation is that the mutual suppression of modes depends on the overlap of three fields: the pump profile and the spatial profiles of the two interacting modes. This leads to the natural conclusion that modes localized in different spatial regions do not suppress each other strongly.

In the absence of relaxation-induced intermode transfer, one can distinguish two regimes: weak competition for $\Delta>0$ and strong competition for $\Delta<0$, where
\[
\Delta=\sigma_{GG}\sigma_{vv}-4\sigma_{Gv}^2.
\]

However, for $\lambda\neq 0$, the mode dynamics is additionally affected by the relaxation-induced coupling coefficient $\rho_{vG}$. Therefore, the sign of $\Delta$ alone does not fully determine the existence and stability of coexistence states, although it remains a useful indicator of the strength of gain-saturation competition. The corresponding criterion is introduced and discussed in the Supplementary Material, Note 1.

The equilibrium points and their stability are analyzed in detail in the Supplementary Material, Note 1. Here we briefly summarize the results. We focus on the regime of strong competition, which is typical under experimental conditions, and restrict ourselves to spatially uniform losses. Then $\tilde \Gamma_G=\tilde \Gamma_v$, which we denote by $\tilde \Gamma$.

If the pump-induced gain coefficient of the ground state exceeds that of the vortex mode $\epsilon_G > \epsilon_v$, the ground state always prevails, and the stationary condensate inherits its symmetry. Pure energy relaxation further favors ground state formation because it introduces additional losses for higher-energy states and gain for the ground state, thereby transferring polaritons from them to the fundamental mode, which has the lowest energy.

A more interesting case arises when the pump profile is such that the vortex mode has a pump-induced gain coefficient larger than that of the fundamental mode, $\epsilon_v>\epsilon_G$. Let us describe the bifurcation scenario that occurs in the absence of energy relaxation as the pump $\mathcal P$ increases.
For sufficiently weak pumping, such that $\epsilon_v \mathcal P < \tilde \Gamma$, the only stationary solution is the trivial state $I_G=I_v=0$, which is globally stable. When the pump $\mathcal P$ reaches the first threshold,
$\mathcal P_{BV}=\frac{\tilde \Gamma}{\epsilon_v},$
a stable vortex-only state,
\[
I_v=\frac{\epsilon_v \mathcal P-\tilde \Gamma}{\sigma_{vv}\mathcal P},\qquad I_G=0,
\]
bifurcates from the trivial solution, while the trivial solution becomes a saddle. We denote this bifurcation by $BV$, which corresponds to the birth of the vortex state.

\begin{figure}[t!]
\includegraphics[width=1.0\linewidth]{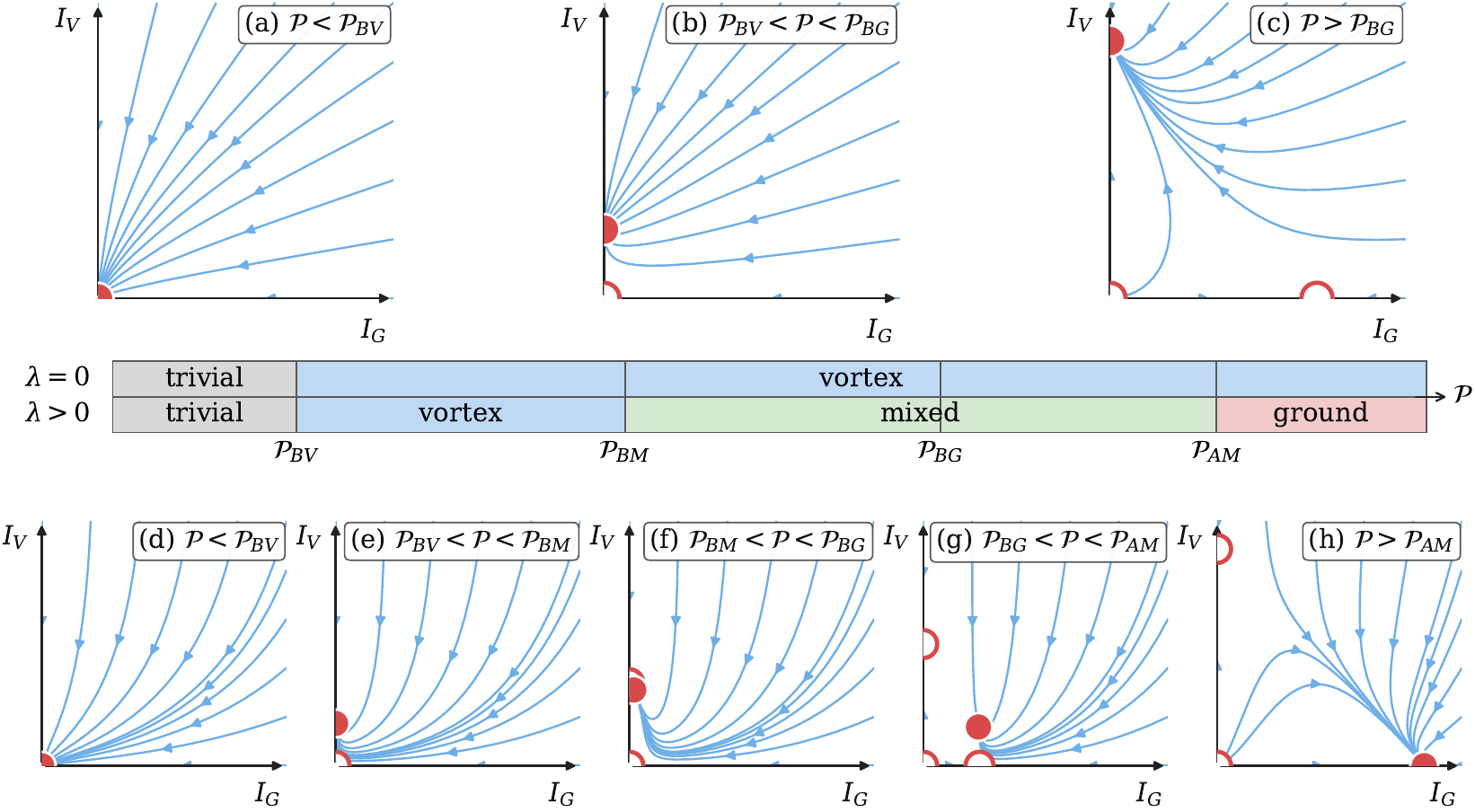}
\caption{Bifurcation scenario for mode competition without (\(\lambda=0\), upper row) and with (\(\lambda>0\), lower row) pure energy relaxation. The central stripe summarizes the asymptotic condensate state as a function of the pump \(\mathcal P\): gray, no condensation (trivial state); blue, vortex state; green, mixed state; red, ground state. The pump thresholds \(\mathcal P_{BV}\), \(\mathcal P_{BM}\), \(\mathcal P_{BG}\), and \(\mathcal P_{AM}\), obtained within the perturbative approach, are indicated on the stripe. Panels (a)--(c) show the phase portraits in the absence of energy relaxation for \(\mathcal P<\mathcal P_{BV}\), \(\mathcal P_{BV}<\mathcal P<\mathcal P_{BG}\), and \(\mathcal P>\mathcal P_{BG}\), respectively. Panels (d)--(h) show the corresponding phase portraits in the presence of energy relaxation for \(\mathcal P<\mathcal P_{BV}\), \(\mathcal P_{BV}<\mathcal P<\mathcal P_{BM}\), \(\mathcal P_{BM}<\mathcal P<\mathcal P_{BG}\), \(\mathcal P_{BG}<\mathcal P<\mathcal P_{AM}\), and \(\mathcal P>\mathcal P_{AM}\), respectively. Filled circles denote stable fixed points, while open circles denote unstable ones.}
\label{bifurcations_chain}
\end{figure}

With further increase of the pump, $\mathcal P>\mathcal P_{BG}=\frac{\tilde\Gamma}{\epsilon_G}$, another equilibrium point bifurcates from the trivial solution. This new solution is the ground-state solution,
\[
I_v=0,\qquad I_G=\frac{\epsilon_G\mathcal P-\tilde\Gamma}{\sigma_{GG}\mathcal P},
\]
which is, however, dynamically unstable (a saddle), while the trivial state becomes an unstable node.

The bifurcation scenario is illustrated in Fig.~\ref{bifurcations_chain}, which shows the thresholds, the pump intervals associated with the formation of different stationary states, and the phase planes for different values of the pump. From this analysis, we conclude that, in the absence of energy relaxation, the dynamics is simple: once the pump exceeds the threshold $\mathcal P_{BV}$, a stable vortex state is formed, and this is the only state realized in the system.

A sufficiently strong pure energy relaxation drastically changes the bifurcation sequence. As the pump crosses the first threshold, a stable vortex state forms (bifurcation $BV$). This state effectively provides an additional gain channel for the ground state. However, for a low-intensity vortex state, this additional gain is not sufficient to make the ground state grow. Thus, only the vortex state can be observed in the system. In the phase plane, there are two equilibrium points: the trivial one (unstable, saddle) and the vortex state (stable node). Therefore, in this pump range, only the vortex state can form, regardless of the presence of the energy-relaxation term.

In the presence of energy relaxation, the excited vortex state facilitates the growth of the ground-state component, and the corresponding bifurcation occurs if the energy-relaxation coefficient is sufficiently large. For smaller values of this coefficient, the depletion of the pump induced by the vortex state cannot be compensated by the transfer of particles from the vortex mode to the ground state.

To determine the threshold, we analyze the stability of the vortex-only state. Linearizing Eq.~\eqref{eq:IG_dyn} with respect to a small ground-state component, we obtain the condition
\begin{equation}
\Gamma_G-
\left(2\sigma_{Gv}\mathcal P-\rho_{vG}\right)
\frac{\Gamma_v}{\sigma_{vv}\mathcal P}=0.
\label{eq:PBM_condition}
\end{equation}
Using $\Gamma_j=-\tilde\Gamma+\epsilon_j\mathcal P$, this condition reduces to
\begin{equation}
(\sigma_{vv}\epsilon_G-2\sigma_{Gv}\epsilon_v)\mathcal P^2
+\Bigl[(2\sigma_{Gv}-\sigma_{vv})\tilde\Gamma+\rho_{vG}\epsilon_v\Bigr]\mathcal P
-\rho_{vG}\tilde\Gamma=0,
\label{eq:PBM_quadratic}
\end{equation}
so that the bifurcation takes place at the positive root of Eq.~\eqref{eq:PBM_quadratic}, which we denote by $\mathcal P_{BM}$.

At $\mathcal P=\mathcal P_{BM}$, a new state bifurcates from the vortex branch, which loses stability and becomes a saddle. The emerging state is linearly stable and corresponds to a superposition of the vortex and ground modes, which have different eigenfrequencies. Within the perturbative description, this state is periodic in time, with period
$T=\frac{2\pi}{\Delta\omega},
\qquad
\Delta\omega=\Omega_v-\Omega_G.$
Thus, the mixed state corresponds to a rotating vortex. We refer to this bifurcation as $BM$, corresponding to the birth of a mixed state.

The next bifurcation occurs when the pump exceeds the threshold $\mathcal P_{BG}$, at which the ground-state solution bifurcates from the trivial state. The ground-state branch is unstable at its onset, while the trivial state becomes an unstable node, as shown in Fig.~\ref{bifurcations_chain}.

To determine the threshold at which the ground-state branch becomes stable, we analyze the stability of the ground-only state.
Linearizing Eq.~\eqref{eq:Iv_dyn} with respect to a small vortex component, we obtain the condition
\begin{equation}
\Gamma_v-
\left(2\sigma_{Gv}\mathcal P+\rho_{vG}\right)
\frac{\Gamma_G}{\sigma_{GG}\mathcal P}=0.
\label{eq:PAM_condition}
\end{equation}
Using $\Gamma_j=-\tilde\Gamma+\epsilon_j\mathcal P$, this condition reduces to
\begin{equation}
(\sigma_{GG}\epsilon_v-2\sigma_{Gv}\epsilon_G)\mathcal P^2
+\Bigl[(2\sigma_{Gv}-\sigma_{GG})\tilde\Gamma-\rho_{vG}\epsilon_G\Bigr]\mathcal P
+\rho_{vG}\tilde\Gamma=0,
\label{eq:PAM_quadratic}
\end{equation}
so that the threshold $\mathcal P_{AM}$ is given by the positive root of Eq.~\eqref{eq:PAM_quadratic}.

At $\mathcal P=\mathcal P_{AM}$, the mixed-state and ground-state branches exchange stability. Above this threshold, the ground state becomes stable, while the mixed state loses stability. We denote this bifurcation by $AM$. In the phase planes, we do not show the unphysical region of negative intensities.

To test these predictions, we solved Eq.~\eqref{GPE} numerically in a deep axisymmetric trap with absorbing boundaries. Numerical simulations were performed in dimensionless units; the conversion to dimensional parameters is given in Supplementary Material, Note 4. The trapping potential was taken in the form
$V_{\mathrm{trap}}(r)=V_0 \frac{1}{1+\exp\!\left(\frac{r-R_0}{h_0}\right)},$
where \mbox{$V_0<0$} is the trap depth, and the pump profile was chosen as
\mbox{$\eta(r)=\eta_0 \left[
\exp\!\left(-\frac{(r-R_1)^2}{h_1^2}\right)
+
\exp\!\left(-\frac{(r+R_1)^2}{h_1^2}\right)\right],$}
with $\eta_0$ fixed by $\max_r \eta(r)=1$.
Low-intensity $\delta$-correlated noise was used as the initial condition, together with a weak random forcing to seed relaxation-induced instabilities. To characterize the condensate, we project the field onto the linear eigenmodes and define
\mbox{$J_m=\left|\int_{\mathbb R^2}\phi_m^*(\mathbf r)\,\psi(\mathbf r,t)\,d^2r\right|^2,$}
where $m=0$ corresponds to the ground state and $m=\pm1$ to the vortex and antivortex modes.

\begin{figure}[t!]
\includegraphics[width=1.0\linewidth]{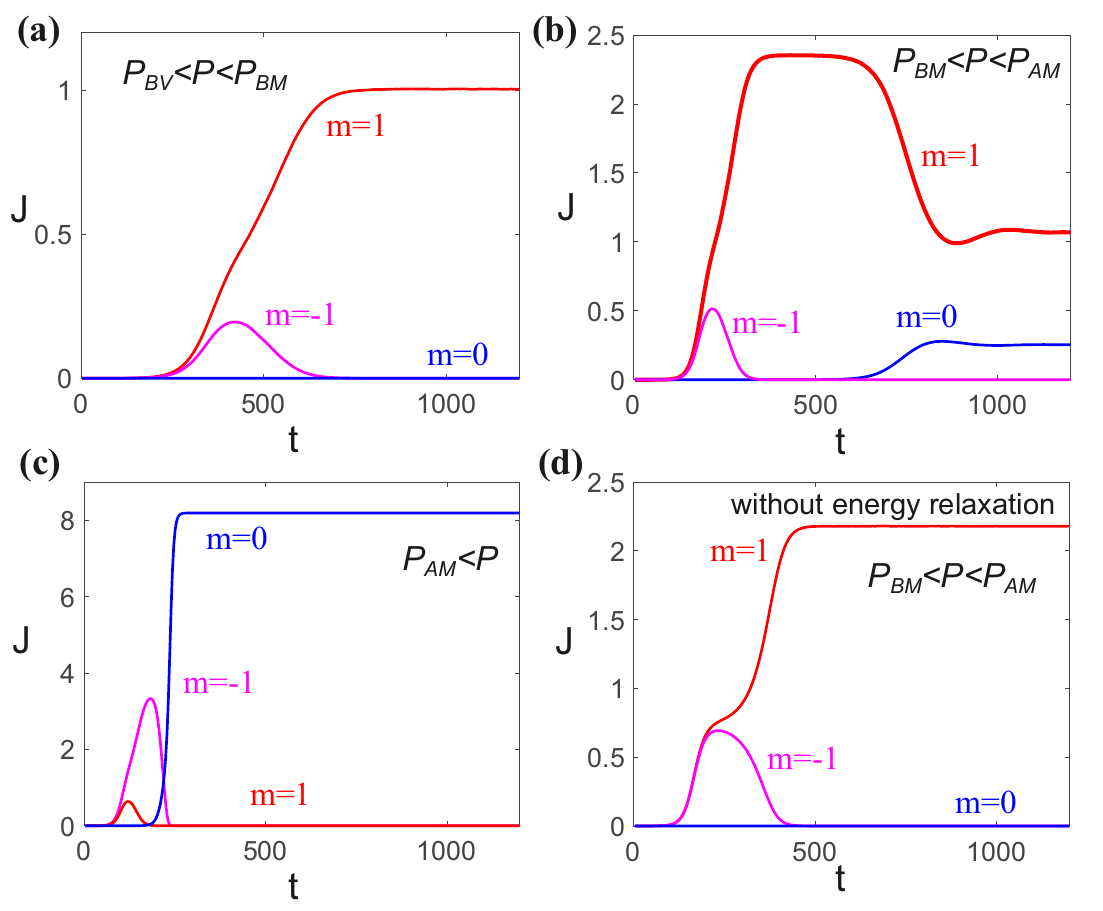}
\caption{Evolution of the modal projections onto the eigenmodes of the linear problem for different pump strengths. Panels (a)--(c) correspond to the case with energy relaxation, while panel (d) shows the same parameters as in panel (b), but without energy relaxation. The blue curves represent the ground state, the red curves the vortex mode with azimuthal index $m=1$, and the magenta curves the vortex mode with azimuthal index $m=-1$. Panels (a)--(c) correspond to pump values ${\cal P}=1.106$, ${\cal P}=1.172$, and ${\cal P}=1.238$, respectively. See the text for details.} \label{formation}
\end{figure}

\begin{figure}[t!]
\includegraphics[width=1.0\linewidth]{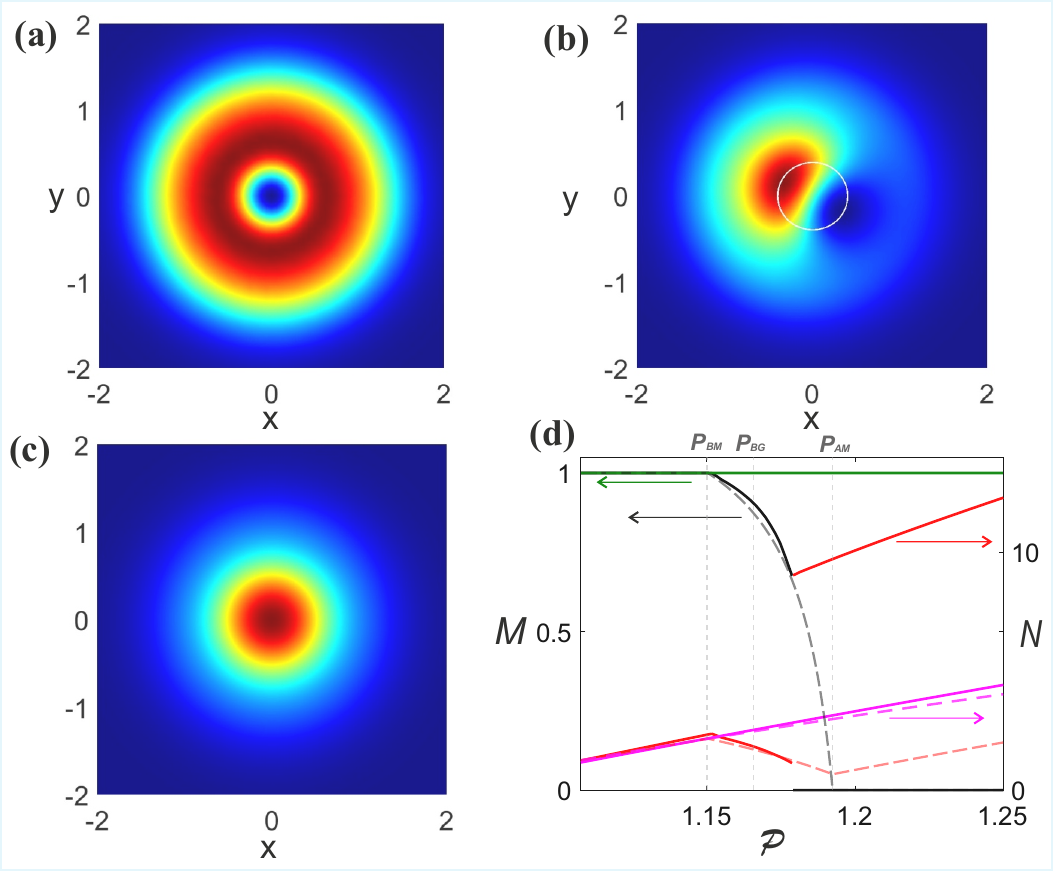}
\caption{Asymptotic condensate states in the presence of energy relaxation. Panels (a)--(c) show the condensate density distributions for three representative pump values. Panel (a) corresponds to the vortex state formed at ${\cal P}=1.106$. Panel (b) shows a snapshot of the rotating mixed state at ${\cal P}=1.172$; the white curve indicates the trajectory of the phase singularity. Panel (c) shows the ground-state condensate formed at ${\cal P}=1.238$. Panel (d) shows the normalized angular momentum ${\cal M}$ versus pump ${\cal P}$ (left axis) and the total condensate population $N$ (right axis), both with and without energy relaxation. Solid curves correspond to numerical simulations of Eq.~\eqref{GPE}, while dashed curves show the predictions of the perturbative theory. The thresholds $\mathcal P_{BM}$, $\mathcal P_{BG}$, and $\mathcal P_{AM}$ are indicated by vertical dashed lines.}
\label{states}
\end{figure}

Let us start with the dynamics in the presence of energy relaxation, $\lambda=0.0025$. Typical dynamics at pump ${\cal P}=1.106$, which is above ${\cal P}_{BV}$ but below ${\cal P}_{BM}$, are shown in Fig.~\ref{formation}(a). As predicted by the perturbative theory, a vortex state forms through competition between the two modes with $m=\pm1$. At early times, both modes with $m=\pm1$ grow exponentially, but at the nonlinear stage their competition results in the survival of a single vortex mode. Figure~\ref{formation}(a) shows the modal occupations obtained from direct numerical simulations of Eq.~\eqref{GPE}, while the corresponding results of the reduced three-mode model ($m=0$, $m=\pm1$) are presented in the Supplementary Material, Note 3. The asymptotic condensate density distribution is shown in Fig.~\ref{states}(a), confirming that the stationary state is an axisymmetric vortex state.

If the pump exceeds the threshold $\mathcal P_{BM}$ but remains below the numerically observed transition to the ground-state regime, the dynamics changes. As seen in Fig.~\ref{formation}(b), the vortex modes with $m=\pm1$ initially grow, after which one of them suppresses its counterpart and a vortex state forms. At longer times, however, the ground state starts to grow, partially suppressing the vortex state. As a result, a rotating mixed state forms; see Fig.~\ref{states}(b), which shows a snapshot of the condensate density distribution. This structure rotates in time with a constant angular velocity; see the movies in the Supplementary Material. The white curve in Fig.~\ref{states}(b) indicates the trajectory of the phase singularity.

Consistent with the perturbative theory, the final state in this regime is determined by whether the pump is below or above $\mathcal P_{AM}$, rather than by $\mathcal P_{BG}$ alone. In particular, for $\mathcal P_{BM}<\mathcal P<\mathcal P_{AM}$ the ground-state component may grow already at intermediate times, but the system ultimately evolves to a rotating mixed state rather than to a pure ground-state condensate.

With further increase of the pump, $ \mathcal P> \mathcal P_{AM}$, we see that, as before, the vortex modes grow faster than the ground state. However, at longer times the dominant vortex mode suppresses the other vortex mode but also triggers the growth of the ground state, which eventually wins the competition; see Fig.~\ref{formation}(c). The density distribution corresponding to the stationary ground state is shown in Fig.~\ref{states}(c).

In the absence of energy relaxation, the dynamics is much simpler. A vortex polariton state appears once the pump exceeds the threshold $\mathcal P_{BV}$. This state forms as a result of the competition between two vortex modes with opposite topological charges. The evolution of the modal occupations for the same pump value as in Fig.~\ref{formation}(b), but without energy relaxation, is shown in Fig.~\ref{formation}(d). If the pump exceeds the threshold $\mathcal P_{BG}$ corresponding to the excitation of the ground state, then the linear growth of the ground state can be observed at the linear stage, but it eventually loses to one of the vortex states.

To quantify the transition between the asymptotic states, we compute the condensate population,
$N=\int |\psi|^2\,dS,$
and the normalized angular-momentum measure,
${\cal M}=\frac{\left| \int \psi^{*}(-i)\left(x\partial_y\psi-y\partial_x\psi\right)\,dx\,dy \right|}{N}.$
The corresponding dependences on the pump are shown in Fig.~\ref{states}(d). Without energy relaxation, ${\cal M}=1$ throughout the considered pump range, indicating a vortex condensate with topological charge $\pm1$, while $N$ grows monotonically with $\mathcal P$.

With energy relaxation, three regimes are observed. At low pump, the condensate is a vortex and ${\cal M}=1$. Above ${\cal P}_{BM}$, ${\cal M}$ decreases continuously as the ground-state fraction grows in the rotating mixed state. Finally, above the numerically observed transition at ${\cal P}_{th}\approx1.18$, the condensate collapses into the ground state, ${\cal M}\to0$, and the population resumes monotonic growth.

The largest discrepancy between the perturbative theory and the simulations is found for the final transition, which is sharper numerically than predicted analytically. This is expected because, for the chosen parameters, the pump-induced blueshift significantly reshapes the underlying modes and thus partially violates the assumptions of the perturbative approach. For smaller $\tilde g$ or a deeper trap, the agreement improves.

In summary, we have shown that pure energy relaxation qualitatively modifies mode selection in nonequilibrium polariton condensation. Within the validity range of the reduced model, it destabilizes vortex condensation at strong pumping, promotes ground-state selection, and gives rise to an intermediate rotating mixed state. These results identify energy relaxation as an essential ingredient of driven-dissipative polariton-condensate dynamics.

\textit{Acknowledgments.} 
This work was supported by the Federal Academic Leadership Program Priority 2030.

\bibliography{main}   

\end{document}


\title{Supplementary Materials: Ground-State Selection by Pure Energy Relaxation in Polariton Condensates}

\author{D. A. Saltykova}
\affiliation{ITMO University, St. Petersburg 197101, Russia}

\author{A. V. Yulin}
\affiliation{ITMO University, St. Petersburg 197101, Russia}

\author{I. A. Shelykh}
\affiliation{Science Institute, University of Iceland, Dunhagi 3, IS-107 Reykjavik, Iceland}

\maketitle
\onecolumngrid
\tableofcontents

\section*{Supplementary Note 1: General equations for two interacting modes}

In this Supplementary Note we formulate the modal reduction in a form that is more general than the one used in the main text. This allows us to present a broader stability and bifurcation analysis, while still making explicit how the equations of the main text are recovered as a particular specialization.

The main text uses orthonormal eigenmodes of the conservative trap, denoted by $\phi_j$, together with modal amplitudes $A_j$ and intensities $I_j=|A_j|^2$. In the present Supplementary Note, we work in a more general basis, denoted by $\varphi_j$, with amplitudes $B_j$ and intensities $\mathcal I_j=|B_j|^2$. This generalized basis may incorporate the pump-induced linear blueshift, but it is normalized in the same $L^2$ sense as the basis used in the main text. Thus, the difference between the two formulations lies in the choice of basis functions rather than in their normalization. The equations of the main text are recovered in the weak-blueshift regime, where the generalized basis $\varphi_j$ reduces to the conservative-trap basis $\phi_j$.

As in the main text, all energies are expressed in frequency units, i.e., divided by $\hbar$.

\subsection*{A. Full condensate--reservoir model}

The driven-dissipative exciton-polariton condensate is described by a generalized Gross--Pitaevskii equation coupled to a rate equation for the incoherent excitonic reservoir. The condensate wavefunction $\psi(\mathbf r,t)$ and the reservoir density $n_R(\mathbf r,t)$ obey
\begin{align}
i \frac{\partial \psi}{\partial t}
&=
\left[
-\frac{\hbar \nabla^2}{2m_{LP}}
+V_{\mathrm{trap}}(r)
+\frac{i}{2}\left(R(n_R)-\gamma\right)
+g|\psi|^2
+2\tilde g\,n_R
\right]\psi
+\frac{i}{2}\lambda \psi \left(\psi^* \nabla^2 \psi-\psi \nabla^2 \psi^*\right),
\label{eq:S1_GPE_full}
\\
\frac{\partial n_R}{\partial t}
&=
P(\mathbf r)-\gamma_R n_R-R(n_R)|\psi|^2+D\nabla^2 n_R .
\label{eq:S1_reservoir_full}
\end{align}

Here $m_{LP}$ is the effective polariton mass, $V_{\mathrm{trap}}(r)$ is the trapping potential, $\gamma$ and $\gamma_R$ are the condensate and reservoir decay rates, $g$ is the polariton--polariton interaction constant, $\tilde g$ is the polariton--reservoir interaction constant, $R(n_R)$ describes stimulated scattering from the reservoir into the condensate, and $D$ is the reservoir diffusion coefficient. The coefficient $\lambda$ describes pure energy relaxation within the condensate.

The terms entering Eq.~\eqref{eq:S1_GPE_full} have the following physical meaning. The operator $-\hbar\nabla^2/(2m_{LP})$ describes the kinetic energy of lower polaritons. The term $V_{\mathrm{trap}}(r)$ represents the external confinement. The imaginary contribution $(i/2)(R(n_R)-\gamma)$ accounts for the gain--loss balance: stimulated feeding from the reservoir competes with intrinsic condensate decay. The cubic term $g|\psi|^2$ describes polariton--polariton interactions, while $2\tilde g\,n_R$ gives the reservoir-induced blueshift. Finally, the term proportional to $\lambda$ describes pure energy relaxation: it conserves particle number but reduces the kinetic energy of the condensate and drives the system toward lower-lying states.

Equation~\eqref{eq:S1_reservoir_full} describes the reservoir dynamics. The pump profile is denoted by $P(\mathbf r)$. The term $-\gamma_R n_R$ accounts for reservoir decay, the term $-R(n_R)|\psi|^2$ describes depletion of the reservoir due to stimulated scattering into the condensate, and $D\nabla^2 n_R$ models spatial transport of excitons.

\subsection*{B. Adiabatic elimination of the reservoir}

We now restrict ourselves to the experimentally relevant fast-reservoir regime, in which $\gamma_R$ is the largest rate in the problem. We also neglect reservoir diffusion and assume a linear scattering law,
\begin{equation}
D=0,
\qquad
R(n_R)=R\,n_R.
\end{equation}
Under these assumptions, the reservoir density adiabatically follows the condensate and is approximately given by
\begin{equation}
n_R(\mathbf r,t)
\simeq
\frac{P(\mathbf r)}{\gamma_R+R|\psi(\mathbf r,t)|^2}.
\label{eq:S1_nR_exact}
\end{equation}
In the weak-saturation regime,
\begin{equation}
R|\psi|^2\ll \gamma_R,
\label{eq:S1_weak_sat}
\end{equation}
we expand Eq.~\eqref{eq:S1_nR_exact} as
\begin{equation}
n_R(\mathbf r,t)
\simeq
\frac{P(\mathbf r)}{\gamma_R}
\left(
1-\frac{R}{\gamma_R}|\psi(\mathbf r,t)|^2
\right).
\label{eq:S1_nR_expanded}
\end{equation}
Substituting this expression into Eq.~\eqref{eq:S1_GPE_full}, we obtain the effective driven-dissipative Gross--Pitaevskii equation in the same form as in the main text,
\begin{equation}
i\frac{\partial \psi}{\partial t}
=
\left[
-\frac{\hbar \nabla^{2}}{2m_{LP}}
+V_{\mathrm{eff}}(|\psi|^{2})
+i\,\Gamma_{\mathrm{eff}}(|\psi|^{2})
\right]\psi
+\frac{i}{2}\lambda \psi \left(\psi^{*}\nabla^{2}\psi-\psi\nabla^{2}\psi^{*}\right),
\label{eq:S1_effective_GPE}
\end{equation}
with
\begin{equation}
V_{\mathrm{eff}}(|\psi|^{2})
=
V_{\mathrm{trap}}(r)
+
g|\psi|^{2}
+
\frac{2\tilde g}{\gamma_R}
\left(
1-\frac{R}{\gamma_R}|\psi|^2
\right)P(\mathbf r),
\label{eq:S1_Veff_general}
\end{equation}
and
\begin{equation}
\Gamma_{\mathrm{eff}}(|\psi|^{2})
=
\frac{1}{2}
\left[
\frac{R}{\gamma_R}
\left(
1-\frac{R}{\gamma_R}|\psi|^2
\right)P(\mathbf r)
-\gamma
\right].
\label{eq:S1_Gammaeff_general}
\end{equation}

For later use, it is convenient to separate the linear conservative contribution from the nonlinear one and rewrite Eq.~\eqref{eq:S1_effective_GPE} as
\begin{equation}
i\partial_t\psi
=
\left[
-\frac{\hbar \nabla^2}{2m_{LP}}
+V_{\mathrm{trap}}(r)
+\frac{2\tilde g\,P(\mathbf r)}{\gamma_R}
\right]\psi
+\bigl(g_{\mathrm{eff}}(\mathbf r)-i\chi(\mathbf r)\bigr)|\psi|^2\psi
+i\Gamma_0(\mathbf r)\psi
+\frac{i\lambda}{2}\psi\left(\psi^*\nabla^2\psi-\psi\nabla^2\psi^*\right),
\label{eq:S1_finalGPE_general}
\end{equation}
where
\begin{align}
g_{\mathrm{eff}}(\mathbf r)
&=
g-\frac{2\tilde g\,R}{\gamma_R^2}P(\mathbf r),
\\
\chi(\mathbf r)
&=
\frac{R^2}{2\gamma_R^2}P(\mathbf r),
\\
\Gamma_0(\mathbf r)
&=
\frac{1}{2}\left(\frac{R\,P(\mathbf r)}{\gamma_R}-\gamma\right).
\end{align}
This form is equivalent to Eq.~\eqref{eq:S1_effective_GPE}, but makes explicit the linear pump-induced blueshift, the effective nonlinear interaction, the nonlinear gain saturation, and the linear gain profile.

\subsection*{C. General modal basis}

In the main text the modal expansion is performed in the orthonormal eigenmodes of the conservative trap. Here we use a more general basis, which may additionally incorporate the pump-induced linear blueshift contained in Eq.~\eqref{eq:S1_finalGPE_general}. Specifically, we define the modes $\varphi_j$ through
\begin{equation}
\left[
-\frac{\hbar \nabla^{2}}{2m_{LP}}
+V_{\mathrm{trap}}(r)
+\frac{2\tilde g\,P(\mathbf r)}{\gamma_R}
\right]\varphi_j(\mathbf r)
=
\omega_j\,\varphi_j(\mathbf r).
\label{eq:S1_general_eigenproblem}
\end{equation}
We focus on the two lowest relevant modes: the fundamental mode and the vortex mode,
\begin{equation}
\varphi_G(r,\theta)=u_G(r),
\qquad
\varphi_v(r,\theta)=u_v(r)e^{i\theta}.
\label{eq:S1_general_modes}
\end{equation}

The generalized modes are normalized in the same way as the modes used in the main text:
\begin{equation}
\int_{\mathbb R^2}\varphi_i^*(\mathbf r)\varphi_j(\mathbf r)\,d^2r=\delta_{ij}.
\label{eq:S1_general_norm}
\end{equation}
For the radial functions introduced in Eq.~\eqref{eq:S1_general_modes}, this implies
\begin{equation}
2\pi\int_0^\infty u_j^2(r)\,r\,dr=1.
\end{equation}

The condensate wavefunction is expanded in this basis as
\begin{equation}
\psi(\mathbf r,t)
=
B_G(t)\,\varphi_G(\mathbf r)\,e^{-i\omega_G t}
+
B_v(t)\,\varphi_v(\mathbf r)\,e^{-i\omega_v t}.
\label{eq:S1_general_ansatz}
\end{equation}
The modal amplitudes are defined by projection:
\begin{equation}
B_j(t)e^{-i\omega_j t}
=
\int_{\mathbb R^2}\varphi_j^*(\mathbf r)\,\psi(\mathbf r,t)\,d^2r,
\qquad
j\in\{G,v\}.
\label{eq:S1_general_projection}
\end{equation}
For a radially symmetric pump profile $P(\mathbf r)=P(r)$, the off-diagonal linear overlaps vanish:
\begin{equation}
\int_{\mathbb R^2}\Gamma_0(r)\,\varphi_G^*(\mathbf r)\varphi_v(\mathbf r)\,d^2r=0.
\end{equation}
Therefore, in the linear approximation the two modes are coupled only through nonlinear saturation and energy relaxation.

\subsection*{D. Relation to the notation of the main text}

The main text uses orthonormal modes $\phi_j$ satisfying
\begin{equation}
\int_{\mathbb R^2}\phi_i^*(\mathbf r)\phi_j(\mathbf r)\,d^2r=\delta_{ij}.
\label{eq:S1_main_orthonorm}
\end{equation}
In the present Supplementary Note, we use the same normalization convention for the generalized modes $\varphi_j$:
\begin{equation}
\int_{\mathbb R^2}\varphi_i^*(\mathbf r)\varphi_j(\mathbf r)\,d^2r=\delta_{ij}.
\end{equation}
Thus, the difference between the two formulations lies only in the choice of the linear basis. The modes $\phi_j$ used in the main text are eigenmodes of the conservative trap, whereas the modes $\varphi_j$ introduced here may incorporate the pump-induced linear blueshift. Therefore, in general,
\[
\varphi_j \neq \phi_j,
\qquad
B_j \neq A_j.
\]

Once the pump is factorized as
\begin{equation}
P(\mathbf r)=\mathcal P\,\eta(r),
\qquad
\max_r \eta(r)=1,
\label{eq:S1_pump_factorization}
\end{equation}
the overlap coefficients can be split into pump-dependent and geometric contributions. In the weak-blueshift regime, where the pump-induced linear blueshift does not substantially reshape the trap eigenmodes, the generalized basis $\varphi_j$ reduces to the conservative-trap basis $\phi_j$, and the generalized amplitudes $B_j$ reduce to the amplitudes $A_j$ used in the main text.

\subsection*{E. Coupled-mode equations for the amplitudes}

Substituting the ansatz \eqref{eq:S1_general_ansatz} into Eq.~\eqref{eq:S1_finalGPE_general}, multiplying from the left by $\varphi_G^* e^{i\omega_G t}$ and integrating over $d^2r$ (and analogously for the vortex mode), we obtain coupled equations for the modal amplitudes. Within the rotating-wave approximation we keep only the resonant contributions and neglect rapidly oscillating terms such as $B_G^2 B_v^* e^{\pm i(\omega_G-\omega_v)t}$.

We define the linear gains
\begin{equation}
\bar\Gamma_j
=
\int_{\mathbb R^2}\Gamma_0(\mathbf r)\,|\varphi_j(\mathbf r)|^2\,d^2r,
\qquad
j\in\{G,v\},
\label{eq:S1_Gammabar}
\end{equation}
the conservative nonlinear coefficients
\begin{align}
\bar u_{GG}
&=
\int_{\mathbb R^2}g_{\mathrm{eff}}(\mathbf r)\,|\varphi_G|^4\,d^2r,
\qquad
\bar u_{vv}
=
\int_{\mathbb R^2}g_{\mathrm{eff}}(\mathbf r)\,|\varphi_v|^4\,d^2r,
\nonumber\\
\bar u_{Gv}
&=
\bar u_{vG}
=
\int_{\mathbb R^2}g_{\mathrm{eff}}(\mathbf r)\,|\varphi_G|^2|\varphi_v|^2\,d^2r,
\label{eq:S1_ubar}
\end{align}
the dissipative saturation coefficients
\begin{align}
\bar\sigma_{GG}
&=
\int_{\mathbb R^2}\chi(\mathbf r)\,|\varphi_G|^4\,d^2r,
\qquad
\bar\sigma_{vv}
=
\int_{\mathbb R^2}\chi(\mathbf r)\,|\varphi_v|^4\,d^2r,
\nonumber\\
\bar\sigma_{Gv}
&=
\bar\sigma_{vG}
=
\int_{\mathbb R^2}\chi(\mathbf r)\,|\varphi_G|^2|\varphi_v|^2\,d^2r,
\label{eq:S1_sigmabar}
\end{align}
and the relaxation-induced intermode transfer coefficient
\begin{equation}
\bar\rho_{vG}
=
\lambda\,\frac{m_{LP}}{\hbar}\,(\omega_v-\omega_G)\,
\int_{\mathbb R^2}|\varphi_G|^2|\varphi_v|^2\,d^2r .
\label{eq:S1_rhobar}
\end{equation}

In terms of these coefficients, the amplitude equations take the form
\begin{align}
i\,\dot{B}_G
&=
i\,\bar\Gamma_G\,B_G
+\bigl(\bar u_{GG}-i\bar\sigma_{GG}\bigr)|B_G|^2 B_G
+2\bigl(\bar u_{Gv}-i\bar\sigma_{Gv}\bigr)|B_v|^2 B_G
+i\,\bar\rho_{vG}|B_v|^2 B_G,
\label{eq:S1_BG_amp}
\\
i\,\dot{B}_v
&=
i\,\bar\Gamma_v\,B_v
+\bigl(\bar u_{vv}-i\bar\sigma_{vv}\bigr)|B_v|^2 B_v
+2\bigl(\bar u_{vG}-i\bar\sigma_{vG}\bigr)|B_G|^2 B_v
-i\,\bar\rho_{vG}|B_G|^2 B_v.
\label{eq:S1_Bv_amp}
\end{align}
The conservative coefficients $\bar u_{ij}$ affect only the phase evolution and nonlinear frequency shifts, whereas the dissipative coefficients $\bar\sigma_{ij}$ and the relaxation coefficient $\bar\rho_{vG}$ determine the intensity dynamics.

\subsection*{F. Amplitude--phase representation and intensity equations}

We write the mode amplitudes in polar form,
\begin{equation}
B_G=R_G e^{i\vartheta_G},
\qquad
B_v=R_v e^{i\vartheta_v},
\end{equation}
where $R_G,R_v\ge 0$ are real amplitudes and $\vartheta_G,\vartheta_v$ are the slow nonlinear phases. Substituting this representation into Eqs.~\eqref{eq:S1_BG_amp} and \eqref{eq:S1_Bv_amp}, we obtain
\begin{align}
\dot R_G
&=
\Big[\bar\Gamma_G-\bar\sigma_{GG}R_G^2-(2\bar\sigma_{Gv}-\bar\rho_{vG})R_v^2\Big]R_G,
\label{eq:S1_RG}
\\
\dot\vartheta_G
&=
-\Big(\bar u_{GG}R_G^2+2\bar u_{Gv}R_v^2\Big),
\label{eq:S1_thetaG}
\\
\dot R_v
&=
\Big[\bar\Gamma_v-\bar\sigma_{vv}R_v^2-(2\bar\sigma_{vG}+\bar\rho_{vG})R_G^2\Big]R_v,
\label{eq:S1_Rv}
\\
\dot\vartheta_v
&=
-\Big(\bar u_{vv}R_v^2+2\bar u_{vG}R_G^2\Big).
\label{eq:S1_thetav}
\end{align}

Introducing the modal intensities
\begin{equation}
\mathcal I_G\equiv R_G^2=|B_G|^2,
\qquad
\mathcal I_v\equiv R_v^2=|B_v|^2,
\end{equation}
we arrive at the autonomous system
\begin{align}
\dot{\mathcal I}_G
&=
2\mathcal I_G
\Big[
\bar\Gamma_G
-\bar\sigma_{GG}\mathcal I_G
-\bigl(2\bar\sigma_{Gv}-\bar\rho_{vG}\bigr)\mathcal I_v
\Big],
\label{eq:S1_IG_general}
\\
\dot{\mathcal I}_v
&=
2\mathcal I_v
\Big[
\bar\Gamma_v
-\bar\sigma_{vv}\mathcal I_v
-\bigl(2\bar\sigma_{vG}+\bar\rho_{vG}\bigr)\mathcal I_G
\Big].
\label{eq:S1_Iv_general}
\end{align}
For brevity, we introduce
\begin{equation}
\bar a=2\bar\sigma_{Gv}-\bar\rho_{vG},
\qquad
\bar b=2\bar\sigma_{vG}+\bar\rho_{vG},
\label{eq:S1_abbar}
\end{equation}
so that the system becomes
\begin{align}
\dot{\mathcal I}_G
&=
2\mathcal I_G\bigl(\bar\Gamma_G-\bar\sigma_{GG}\mathcal I_G-\bar a\,\mathcal I_v\bigr),
\label{eq:S1_IG_compact}
\\
\dot{\mathcal I}_v
&=
2\mathcal I_v\bigl(\bar\Gamma_v-\bar\sigma_{vv}\mathcal I_v-\bar b\,\mathcal I_G\bigr).
\label{eq:S1_Iv_compact}
\end{align}

\subsection*{G. Fixed points and Jacobian}

The fixed points of the system \eqref{eq:S1_IG_compact}--\eqref{eq:S1_Iv_compact} satisfy
\[
\dot{\mathcal I}_G=0,
\qquad
\dot{\mathcal I}_v=0.
\]
There are up to four candidate fixed points in the physically admissible quadrant $\mathcal I_G\ge 0$, $\mathcal I_v\ge 0$.

Define
\[
\mathbf F(\mathcal I_G,\mathcal I_v)
=
\begin{pmatrix}
f_G(\mathcal I_G,\mathcal I_v)\\
f_v(\mathcal I_G,\mathcal I_v)
\end{pmatrix}
=
\begin{pmatrix}
2\mathcal I_G(\bar\Gamma_G-\bar\sigma_{GG}\mathcal I_G-\bar a\,\mathcal I_v)\\
2\mathcal I_v(\bar\Gamma_v-\bar\sigma_{vv}\mathcal I_v-\bar b\,\mathcal I_G)
\end{pmatrix}.
\]
Then the Jacobian is
\begin{equation}
J(\mathcal I_G,\mathcal I_v)
=
2
\begin{pmatrix}
\bar\Gamma_G-2\bar\sigma_{GG}\mathcal I_G-\bar a\,\mathcal I_v
&
-\bar a\,\mathcal I_G
\\[4pt]
-\bar b\,\mathcal I_v
&
\bar\Gamma_v-2\bar\sigma_{vv}\mathcal I_v-\bar b\,\mathcal I_G
\end{pmatrix}.
\label{eq:S1_Jacobian}
\end{equation}

The four candidate fixed points are as follows.

\noindent\textbf{(i) Trivial state.}
\begin{equation}
(\mathcal I_G^0,\mathcal I_v^0)=(0,0).
\end{equation}
At this point,
\begin{equation}
J(0,0)
=
2
\begin{pmatrix}
\bar\Gamma_G & 0\\
0 & \bar\Gamma_v
\end{pmatrix},
\end{equation}
and the eigenvalues are
\begin{equation}
\Lambda_1=2\bar\Gamma_G,
\qquad
\Lambda_2=2\bar\Gamma_v.
\end{equation}

\noindent\textbf{(ii) Ground-only state.}
\begin{equation}
(\mathcal I_G^0,\mathcal I_v^0)
=
\left(\frac{\bar\Gamma_G}{\bar\sigma_{GG}},\,0\right),
\qquad
\bar\Gamma_G>0.
\end{equation}
The Jacobian becomes
\begin{equation}
J\left(\frac{\bar\Gamma_G}{\bar\sigma_{GG}},0\right)
=
2
\begin{pmatrix}
-\bar\Gamma_G &
-\bar a\,\dfrac{\bar\Gamma_G}{\bar\sigma_{GG}}
\\[6pt]
0 &
\bar\Gamma_v-\bar b\,\dfrac{\bar\Gamma_G}{\bar\sigma_{GG}}
\end{pmatrix},
\end{equation}
hence
\begin{equation}
\Lambda_1=-2\bar\Gamma_G,
\qquad
\Lambda_2=
2\left(
\bar\Gamma_v-\frac{\bar b}{\bar\sigma_{GG}}\bar\Gamma_G
\right).
\label{eq:S1_ground_transverse}
\end{equation}

\noindent\textbf{(iii) Vortex-only state.}
\begin{equation}
(\mathcal I_G^0,\mathcal I_v^0)
=
\left(0,\,\frac{\bar\Gamma_v}{\bar\sigma_{vv}}\right),
\qquad
\bar\Gamma_v>0.
\end{equation}

Its Jacobian is
\begin{equation}
J\left(0,\frac{\bar\Gamma_v}{\bar\sigma_{vv}}\right)
=
2
\begin{pmatrix}
\bar\Gamma_G-\bar a\,\dfrac{\bar\Gamma_v}{\bar\sigma_{vv}} & 0
\\[6pt]
-\bar b\,\dfrac{\bar\Gamma_v}{\bar\sigma_{vv}} & -\bar\Gamma_v
\end{pmatrix},
\end{equation}
hence
\begin{equation}
\Lambda_1=-2\bar\Gamma_v,
\qquad
\Lambda_2=
2\left(
\bar\Gamma_G-\frac{\bar a}{\bar\sigma_{vv}}\bar\Gamma_v
\right).
\label{eq:S1_vortex_transverse}
\end{equation}

\noindent\textbf{(iv) Mixed two-mode state.}
Solving the stationary equations
\begin{align}
\bar\Gamma_G-\bar\sigma_{GG}\mathcal I_G^0-\bar a\,\mathcal I_v^0&=0,
\label{eq:S1_mixed_stat1}
\\
\bar\Gamma_v-\bar\sigma_{vv}\mathcal I_v^0-\bar b\,\mathcal I_G^0&=0,
\label{eq:S1_mixed_stat2}
\end{align}
we obtain
\begin{align}
\mathcal I_G^0
&=
\frac{\bar\Gamma_G\bar\sigma_{vv}-\bar a\,\bar\Gamma_v}{\bar\Delta},
\\
\mathcal I_v^0
&=
\frac{\bar\Gamma_v\bar\sigma_{GG}-\bar b\,\bar\Gamma_G}{\bar\Delta},
\end{align}
where
\begin{equation}
\bar\Delta
=
\bar\sigma_{GG}\bar\sigma_{vv}-\bar a\bar b.
\label{eq:S1_Deltabar}
\end{equation}

Using Eqs.~\eqref{eq:S1_mixed_stat1} and \eqref{eq:S1_mixed_stat2}, the Jacobian at the mixed state simplifies to
\begin{equation}
J^0
=
2
\begin{pmatrix}
-\bar\sigma_{GG}\mathcal I_G^0 & -\bar a\,\mathcal I_G^0
\\[6pt]
-\bar b\,\mathcal I_v^0 & -\bar\sigma_{vv}\mathcal I_v^0
\end{pmatrix}.
\end{equation}
Therefore,
\begin{align}
\mathrm{Tr}\,J^0
&=
-2\bigl(\bar\sigma_{GG}\mathcal I_G^0+\bar\sigma_{vv}\mathcal I_v^0\bigr),
\\[4pt]
\det J^0
&=
4\,\mathcal I_G^0\mathcal I_v^0\,\bar\Delta.
\label{eq:S1_TrDet_general}
\end{align}
Hence, whenever the mixed state lies in the physical quadrant:
\begin{itemize}
\item if $\bar\Delta>0$, it is linearly stable;
\item if $\bar\Delta<0$, it is necessarily a saddle;
\item if $\bar\Delta=0$, it is non-hyperbolic.
\end{itemize}

\subsection*{H. Phase portrait for zero relaxation ($\lambda=0$)}

We first consider the case $\lambda=0$, which implies
\begin{equation}
\bar\rho_{vG}=0
\qquad \Longrightarrow \qquad
\bar a=\bar b=2\bar\sigma_{Gv}.
\end{equation}
The intensity equations reduce to
\begin{align}
\dot{\mathcal I}_G
&=
2\mathcal I_G\bigl(\bar\Gamma_G-\bar\sigma_{GG}\mathcal I_G-2\bar\sigma_{Gv}\mathcal I_v\bigr),
\label{eq:S1_IG_lambda0}
\\
\dot{\mathcal I}_v
&=
2\mathcal I_v\bigl(\bar\Gamma_v-\bar\sigma_{vv}\mathcal I_v-2\bar\sigma_{Gv}\mathcal I_G\bigr).
\label{eq:S1_Iv_lambda0}
\end{align}

\paragraph{\bfseries Case $\bar\Gamma_G=\bar\Gamma_v=0$.}
For vanishing linear gains, Eqs.~\eqref{eq:S1_IG_lambda0} and \eqref{eq:S1_Iv_lambda0} become
\begin{equation}
\dot{\mathcal I}_G
=
-2\mathcal I_G\bigl(\bar\sigma_{GG}\mathcal I_G+2\bar\sigma_{Gv}\mathcal I_v\bigr),
\qquad
\dot{\mathcal I}_v
=
-2\mathcal I_v\bigl(\bar\sigma_{vv}\mathcal I_v+2\bar\sigma_{Gv}\mathcal I_G\bigr).
\end{equation}
In the first quadrant, the trivial state is the only equilibrium, and all trajectories monotonically approach it. Thus, the trivial state is globally asymptotically stable in the physical region.

\paragraph{\bfseries Case $\bar\Gamma_G>0$, $\bar\Gamma_v=0$.}
The fixed points are the trivial state and the ground-only state,
\begin{equation}
(\mathcal I_G^0,\mathcal I_v^0)=(0,0),
\qquad
(\mathcal I_G^0,\mathcal I_v^0)=\left(\frac{\bar\Gamma_G}{\bar\sigma_{GG}},\,0\right).
\end{equation}
The ground-only state is asymptotically stable for $\bar\sigma_{Gv}>0$, while the trivial state is unstable along the $\mathcal I_G$ direction.

\paragraph{\bfseries Case $\bar\Gamma_G=0$, $\bar\Gamma_v>0$.}
This case is completely analogous upon exchanging the roles of the two modes. The vortex-only state
\[
\left(0,\frac{\bar\Gamma_v}{\bar\sigma_{vv}}\right)
\]
is asymptotically stable for $\bar\sigma_{Gv}>0$.

\paragraph{\bfseries Weak-competition regime: $\bar\Delta>0$.}
Assume in the following comparison that $\bar\Gamma_G>0$, so that the ground-only branch exists. In the absence of relaxation,
\begin{equation}
\bar\Delta
=
\bar\sigma_{GG}\bar\sigma_{vv}-4\bar\sigma_{Gv}^2.
\end{equation}
Define
\begin{equation}
\bar\Gamma_v^{(1)}
=
\frac{2\bar\sigma_{Gv}}{\bar\sigma_{GG}}\bar\Gamma_G,
\qquad
\bar\Gamma_v^{(2)}
=
\frac{\bar\sigma_{vv}}{2\bar\sigma_{Gv}}\bar\Gamma_G.
\end{equation}
For $\bar\Delta>0$, these thresholds satisfy
\begin{equation}
\bar\Gamma_v^{(1)}<\bar\Gamma_v^{(2)}.
\end{equation}
The mixed state exists in the first quadrant if and only if
\begin{equation}
\bar\Gamma_v^{(1)}<\bar\Gamma_v<\bar\Gamma_v^{(2)}.
\label{eq:S1_mixed_exist_weak}
\end{equation}
Whenever it exists, it is linearly stable. Thus:
\begin{itemize}
\item for $0\le \bar\Gamma_v<\bar\Gamma_v^{(1)}$, the ground-only state is the attractor;
\item for $\bar\Gamma_v^{(1)}<\bar\Gamma_v<\bar\Gamma_v^{(2)}$, the mixed state is the attractor;
\item for $\bar\Gamma_v>\bar\Gamma_v^{(2)}$, the vortex-only state is the attractor.
\end{itemize}
The exchange of stability occurs through two transcritical bifurcations.

\paragraph{\bfseries Strong-competition regime: $\bar\Delta<0$.}
Assume again that $\bar\Gamma_G>0$. If
\begin{equation}
\bar\Delta
=
\bar\sigma_{GG}\bar\sigma_{vv}-4\bar\sigma_{Gv}^2<0,
\end{equation}
the mixed state is always a saddle whenever it exists algebraically. In this case
\begin{equation}
\bar\Gamma_v^{(2)}<\bar\Gamma_v^{(1)},
\end{equation}
and there is a bistability window
\begin{equation}
\bar\Gamma_v^{(2)}<\bar\Gamma_v<\bar\Gamma_v^{(1)},
\end{equation}
within which both single-mode states are stable and the mixed saddle separates their basins of attraction. Outside this interval the attractor is unique.

\begin{table}[t]
\centering
\caption{Classification of fixed points in the first quadrant ($\mathcal I_G\ge0$, $\mathcal I_v\ge0$) for the $\lambda=0$ system.}
\label{tab:S1_fp_classification_lambda0}
\begin{tabular}{p{3.7cm} p{4.7cm} p{6.5cm}}
\hline
Fixed point & Existence (in $\mathcal I_G,\mathcal I_v\ge0$) & Type / stability conditions \\ \hline

Trivial $(0,0)$ &
always &
Hyperbolic if $\bar\Gamma_G\bar\Gamma_v\neq 0$: stable node for $\bar\Gamma_G<0,\bar\Gamma_v<0$; unstable node for $\bar\Gamma_G>0,\bar\Gamma_v>0$; saddle for $\bar\Gamma_G\bar\Gamma_v<0$.  
Non-hyperbolic if $\bar\Gamma_G=0$ or $\bar\Gamma_v=0$ (linearization is then inconclusive; for $\bar\Gamma_G=\bar\Gamma_v=0$ the trivial state is globally asymptotically stable in the first quadrant). \\[6pt]

Ground-only $\bigl(\bar\Gamma_G/\bar\sigma_{GG},0\bigr)$ &
$\bar\Gamma_G>0$ &
Stable node if $\bar\Gamma_v<\bar\Gamma_v^{(1)}$; saddle if $\bar\Gamma_v>\bar\Gamma_v^{(1)}$; non-hyperbolic at $\bar\Gamma_v=\bar\Gamma_v^{(1)}$ (transcritical point). \\[6pt]

Vortex-only $\bigl(0,\bar\Gamma_v/\bar\sigma_{vv}\bigr)$ &
$\bar\Gamma_v>0$ &
Stable node if $\bar\Gamma_v>\bar\Gamma_v^{(2)}$; saddle if $\bar\Gamma_v<\bar\Gamma_v^{(2)}$; non-hyperbolic at $\bar\Gamma_v=\bar\Gamma_v^{(2)}$ (transcritical point). \\[6pt]

Mixed $(\mathcal I_G^0,\mathcal I_v^0)$ &
$\mathcal I_G^0>0,\ \mathcal I_v^0>0$ (when it exists) &
If $\bar\Delta>0$: stable node.  
If $\bar\Delta<0$: saddle.  
Degenerate at $\bar\Delta=0$ (non-hyperbolic). \\ \hline
\end{tabular}
\end{table}

For convenience, in Table~\ref{tab:S1_fp_classification_lambda0} we used
\[
\bar\Delta=\bar\sigma_{GG}\bar\sigma_{vv}-4\bar\sigma_{Gv}^2,
\qquad
\bar\Gamma_v^{(1)}=\frac{2\bar\sigma_{Gv}}{\bar\sigma_{GG}}\bar\Gamma_G,
\qquad
\bar\Gamma_v^{(2)}=\frac{\bar\sigma_{vv}}{2\bar\sigma_{Gv}}\bar\Gamma_G.
\]

\subsection*{I. Phase portrait for finite relaxation ($\lambda\neq 0$)}

For $\lambda\neq 0$, the relaxation-induced coupling
\begin{equation}
\bar\rho_{vG}\propto \lambda(\omega_v-\omega_G)
\end{equation}
is generally nonzero. In the physically relevant case $\lambda>0$ and $\omega_v>\omega_G$, one has $\bar\rho_{vG}>0$. The effective cross-saturation coefficients become asymmetric:
\begin{equation}
\bar a=2\bar\sigma_{Gv}-\bar\rho_{vG},
\qquad
\bar b=2\bar\sigma_{Gv}+\bar\rho_{vG}.
\label{eq:S1_ab_general_relax}
\end{equation}

The mixed-state coordinates are
\begin{equation}
\mathcal I_G^0
=
\frac{\bar\Gamma_G\bar\sigma_{vv}-\bar a\,\bar\Gamma_v}{\bar\Delta},
\qquad
\mathcal I_v^0
=
\frac{\bar\Gamma_v\bar\sigma_{GG}-\bar b\,\bar\Gamma_G}{\bar\Delta},
\end{equation}
with
\begin{equation}
\bar\Delta
=
\bar\sigma_{GG}\bar\sigma_{vv}-\bar a\bar b.
\end{equation}
Using Eq.~\eqref{eq:S1_ab_general_relax}, we obtain
\begin{equation}
\bar a\bar b
=
(2\bar\sigma_{Gv}-\bar\rho_{vG})(2\bar\sigma_{Gv}+\bar\rho_{vG})
=
4\bar\sigma_{Gv}^2-\bar\rho_{vG}^2,
\end{equation}
hence
\begin{equation}
\bar\Delta
=
\bar\sigma_{GG}\bar\sigma_{vv}-4\bar\sigma_{Gv}^2+\bar\rho_{vG}^2.
\label{eq:S1_Delta_relax}
\end{equation}
Thus, increasing the relaxation strength tends to increase $\bar\Delta$ and may suppress the strong-competition scenario present at $\lambda=0$.

\paragraph*{Transverse stability of the one-mode branches.}
At the ground-only branch, the transverse eigenvalue is
\begin{equation}
\Lambda_{\perp}^{(G)}
=
2\left(
\bar\Gamma_v-\frac{\bar b}{\bar\sigma_{GG}}\bar\Gamma_G
\right).
\end{equation}
Therefore, the ground-only state is transversely stable when
\begin{equation}
\bar\Gamma_v<\bar\Gamma_v^{(1)},
\qquad
\bar\Gamma_v^{(1)}\equiv \frac{\bar b}{\bar\sigma_{GG}}\bar\Gamma_G.
\label{eq:S1_gamma1_relax}
\end{equation}

At the vortex-only branch, the transverse eigenvalue is
\begin{equation}
\Lambda_{\perp}^{(v)}
=
2\left(
\bar\Gamma_G-\frac{\bar a}{\bar\sigma_{vv}}\bar\Gamma_v
\right).
\end{equation}
For $\bar a>0$, the vortex-only state is transversely stable when
\begin{equation}
\bar\Gamma_v>\bar\Gamma_v^{(2)},
\qquad
\bar\Gamma_v^{(2)}\equiv \frac{\bar\sigma_{vv}}{\bar a}\bar\Gamma_G.
\label{eq:S1_gamma2_relax}
\end{equation}

\paragraph*{Ordering of thresholds for $\bar a>0$.}
Assume in the following comparison that $\bar\Gamma_G>0$, so that the ground-only branch exists. If $\bar a>0$ and $\bar b>0$, the ordering of the two thresholds is controlled by the sign of $\bar\Delta$:
\begin{equation}
\bar\Gamma_v^{(1)}<\bar\Gamma_v^{(2)}
\quad \Longleftrightarrow \quad
\bar\Delta>0.
\end{equation}
Hence:
\begin{itemize}
\item if $\bar\Delta>0$, the mixed state is stable whenever it exists in the first quadrant;
\item if $\bar\Delta<0$, the mixed state is a saddle and a bistability window appears.
\end{itemize}

\paragraph*{The case $\bar a\le 0$.}
A qualitatively new regime arises if the relaxation is sufficiently strong that
\begin{equation}
\bar a=2\bar\sigma_{Gv}-\bar\rho_{vG}\le 0.
\label{eq:S1_a_negative}
\end{equation}
In this regime the vortex intensity no longer suppresses the ground mode; instead, it tends to enhance its growth.

If $\bar a=0$, the ground-mode equation becomes independent of $\mathcal I_v$:
\begin{equation}
\dot{\mathcal I}_G
=
2\mathcal I_G(\bar\Gamma_G-\bar\sigma_{GG}\mathcal I_G).
\end{equation}
Hence the ground-only branch exists for $\bar\Gamma_G>0$, while the vortex-only branch is transversely stable for $\bar\Gamma_G<0$, loses stability at $\bar\Gamma_G=0$, and is unstable for $\bar\Gamma_G>0$. For $\bar\Gamma_G>0$, a mixed steady state exists when
\begin{equation}
\bar\Gamma_v>\bar\Gamma_v^{(1)},
\qquad
\bar\Gamma_v^{(1)}=\frac{\bar b}{\bar\sigma_{GG}}\bar\Gamma_G,
\end{equation}
and is linearly stable whenever it lies in the first quadrant.

If $\bar a<0$, the transverse eigenvalue of the vortex-only branch is
\begin{equation}
\Lambda_{\perp}^{(v)}
=
2\left(
\bar\Gamma_G+\frac{|\bar a|}{\bar\sigma_{vv}}\bar\Gamma_v
\right).
\label{eq:S1_vortex_transverse_a_negative}
\end{equation}
Therefore, the vortex-only state is transversely stable when
\begin{equation}
\bar\Gamma_G+\frac{|\bar a|}{\bar\sigma_{vv}}\bar\Gamma_v<0,
\label{eq:S1_vortex_stability_a_negative}
\end{equation}
non-hyperbolic at equality, and unstable otherwise.

Since $\bar a<0$ and $\bar b>0$, one has
\begin{equation}
\bar\Delta=\bar\sigma_{GG}\bar\sigma_{vv}-\bar a\bar b>0,
\end{equation}
so the mixed state is linearly stable whenever it exists in the first quadrant. In particular:
\begin{itemize}
\item if $\bar\Gamma_G<0$, the vortex-only state is stable for
\[
0<\bar\Gamma_v<-\frac{\bar\sigma_{vv}}{|\bar a|}\bar\Gamma_G,
\]
and a stable mixed state appears for larger $\bar\Gamma_v$;
\item if $\bar\Gamma_G>0$, the vortex-only branch is always transversely unstable, while the ground-only state is stable for $\bar\Gamma_v<\bar\Gamma_v^{(1)}$ and the mixed state is stable for $\bar\Gamma_v>\bar\Gamma_v^{(1)}$.
\end{itemize}

\begin{table}[t]
\centering
\caption{Classification of fixed points in the first quadrant ($\mathcal I_G\ge0$, $\mathcal I_v\ge0$) for finite relaxation, $\lambda\neq0$.}
\label{tab:S1_fp_classification_relax}
\begin{tabular}{p{3.7cm} p{4.7cm} p{6.5cm}}
\hline
Fixed point & Existence (in $\mathcal I_G,\mathcal I_v\ge0$) & Type / stability conditions \\ \hline

Trivial $(0,0)$ &
always &
Hyperbolic if $\bar\Gamma_G\bar\Gamma_v\neq 0$: stable node for $\bar\Gamma_G<0,\bar\Gamma_v<0$; unstable node for $\bar\Gamma_G>0,\bar\Gamma_v>0$; saddle for $\bar\Gamma_G\bar\Gamma_v<0$.  
Non-hyperbolic if $\bar\Gamma_G=0$ or $\bar\Gamma_v=0$. \\[6pt]

Ground-only $\bigl(\bar\Gamma_G/\bar\sigma_{GG},0\bigr)$ &
$\bar\Gamma_G>0$ &
Stable if $\bar\Gamma_v<\bar\Gamma_v^{(1)}=(\bar b/\bar\sigma_{GG})\bar\Gamma_G$; unstable if $\bar\Gamma_v>\bar\Gamma_v^{(1)}$; non-hyperbolic at $\bar\Gamma_v=\bar\Gamma_v^{(1)}$. \\[6pt]

Vortex-only $\bigl(0,\bar\Gamma_v/\bar\sigma_{vv}\bigr)$ &
$\bar\Gamma_v>0$ &
For $\bar a>0$: stable if $\bar\Gamma_v>\bar\Gamma_v^{(2)}=(\bar\sigma_{vv}/\bar a)\bar\Gamma_G$, unstable if $\bar\Gamma_v<\bar\Gamma_v^{(2)}$, and non-hyperbolic at $\bar\Gamma_v=\bar\Gamma_v^{(2)}$.  
For $\bar a=0$: transversely stable if $\bar\Gamma_G<0$, unstable if $\bar\Gamma_G>0$, and non-hyperbolic at $\bar\Gamma_G=0$.  
For $\bar a<0$: transversely stable if $\bar\Gamma_G+(|\bar a|/\bar\sigma_{vv})\bar\Gamma_v<0$, unstable if $\bar\Gamma_G+(|\bar a|/\bar\sigma_{vv})\bar\Gamma_v>0$, and non-hyperbolic at equality. \\[6pt]

Mixed $(\mathcal I_G^0,\mathcal I_v^0)$ &
$\mathcal I_G^0\ge0,\ \mathcal I_v^0\ge0$, with
$\mathcal I_G^0=(\bar\Gamma_G\bar\sigma_{vv}-\bar a\bar\Gamma_v)/\bar\Delta$ and
$\mathcal I_v^0=(\bar\Gamma_v\bar\sigma_{GG}-\bar b\bar\Gamma_G)/\bar\Delta$ &
If $\bar\Delta>0$: linearly stable whenever it exists in the first quadrant.  
If $\bar\Delta<0$: saddle whenever it exists.  
If $\bar\Delta=0$: non-hyperbolic.  
For $\bar a<0$, the mixed state may be either a stable node or a stable focus. \\ \hline
\end{tabular}
\end{table}

For convenience, in Table~\ref{tab:S1_fp_classification_relax} we used
\[
\bar a=2\bar\sigma_{Gv}-\bar\rho_{vG},
\qquad
\bar b=2\bar\sigma_{Gv}+\bar\rho_{vG},
\qquad
\bar\Delta=\bar\sigma_{GG}\bar\sigma_{vv}-\bar a\bar b.
\]

\subsection*{J. Recovery of the equations used in the main text}

We now show explicitly how the equations used in the main text follow from the more general formulation developed above.

First, the pump is factorized as
\begin{equation}
P(\mathbf r)=\mathcal P\,\eta(r),
\qquad
\max_r \eta(r)=1.
\end{equation}
Second, one assumes that the pump-induced linear blueshift does not substantially reshape the underlying trap modes. In that weak-blueshift regime, the generalized basis $\varphi_j$ reduces to the orthonormal conservative-trap eigenmodes $\phi_j$ of the main text, and the amplitudes $B_j$ reduce to the amplitudes $A_j$ used there:
\begin{equation}
\left[
-\frac{\hbar \nabla^{2}}{2m_{LP}}
+V_{\mathrm{trap}}(r)
\right]\phi_j(\mathbf r)
=
\Omega_j\phi_j(\mathbf r),
\qquad
\int_{\mathbb R^2}\phi_i^*(\mathbf r)\phi_j(\mathbf r)\,d^2r=\delta_{ij}.
\end{equation}

In this limit, the effective linear growth rates become
\begin{equation}
\Gamma_j=-\tilde\Gamma_j+\epsilon_j\mathcal P,
\qquad
j\in\{G,v\},
\end{equation}
where
\begin{equation}
\tilde\Gamma_j
=
\frac{1}{2}
\int_{\mathbb R^2}\gamma\,|\phi_j(\mathbf r)|^2\,d^2r,
\qquad
\epsilon_j
=
\frac{R}{2\gamma_R}
\int_{\mathbb R^2}\eta(r)\,|\phi_j(\mathbf r)|^2\,d^2r.
\end{equation}
The nonlinear gain-saturation coefficients are
\begin{equation}
\sigma_{jj}
=
\frac{R^2}{2\gamma_R^2}
\int_{\mathbb R^2}\eta(r)\,|\phi_j(\mathbf r)|^4\,d^2r,
\qquad
\sigma_{Gv}
=
\frac{R^2}{2\gamma_R^2}
\int_{\mathbb R^2}\eta(r)\,|\phi_G(\mathbf r)|^2|\phi_v(\mathbf r)|^2\,d^2r,
\end{equation}
and the relaxation coefficient is
\begin{equation}
\rho_{vG}
=
\lambda\,\frac{m_{LP}}{\hbar}\,(\Omega_v-\Omega_G)
\int_{\mathbb R^2}|\phi_G(\mathbf r)|^2|\phi_v(\mathbf r)|^2\,d^2r.
\end{equation}

The intensity equations then reduce to
\begin{align}
\dot I_G
&=
2I_G\Bigl(\Gamma_G-\sigma_{GG}\mathcal P\,I_G-\bigl(2\sigma_{Gv}\mathcal P-\rho_{vG}\bigr)I_v\Bigr),
\\
\dot I_v
&=
2I_v\Bigl(\Gamma_v-\sigma_{vv}\mathcal P\,I_v-\bigl(2\sigma_{Gv}\mathcal P+\rho_{vG}\bigr)I_G\Bigr),
\end{align}
which are precisely the equations used in the main text.

For spatially uniform losses, one has
\begin{equation}
\tilde\Gamma_G=\tilde\Gamma_v\equiv \tilde\Gamma,
\end{equation}
so that
\begin{equation}
\Gamma_G=-\tilde\Gamma+\epsilon_G\mathcal P,
\qquad
\Gamma_v=-\tilde\Gamma+\epsilon_v\mathcal P.
\end{equation}
The one-mode condensation thresholds are therefore
\begin{equation}
\mathcal P_{BG}=\frac{\tilde\Gamma}{\epsilon_G},
\qquad
\mathcal P_{BV}=\frac{\tilde\Gamma}{\epsilon_v}.
\end{equation}

The threshold $\mathcal P_{BM}$ at which the vortex-only state loses stability is obtained from the condition
\begin{equation}
\Gamma_G-
\left(2\sigma_{Gv}\mathcal P-\rho_{vG}\right)
\frac{\Gamma_v}{\sigma_{vv}\mathcal P}=0,
\label{eq:S1_PBM_condition}
\end{equation}
which reduces to
\begin{equation}
(\sigma_{vv}\epsilon_G-2\sigma_{Gv}\epsilon_v)\mathcal P^2
+\Bigl[(2\sigma_{Gv}-\sigma_{vv})\tilde\Gamma+\rho_{vG}\epsilon_v\Bigr]\mathcal P
-\rho_{vG}\tilde\Gamma=0.
\label{eq:S1_PBM_quadratic}
\end{equation}
The physically relevant threshold $\mathcal P_{BM}$ is the positive root of Eq.~\eqref{eq:S1_PBM_quadratic}.

Similarly, the threshold $\mathcal P_{AM}$ at which the ground-only state becomes stable is obtained from
\begin{equation}
\Gamma_v-
\left(2\sigma_{Gv}\mathcal P+\rho_{vG}\right)
\frac{\Gamma_G}{\sigma_{GG}\mathcal P}=0,
\label{eq:S1_PAM_condition}
\end{equation}
which reduces to
\begin{equation}
(\sigma_{GG}\epsilon_v-2\sigma_{Gv}\epsilon_G)\mathcal P^2
+\Bigl[(2\sigma_{Gv}-\sigma_{GG})\tilde\Gamma-\rho_{vG}\epsilon_G\Bigr]\mathcal P
+\rho_{vG}\tilde\Gamma=0.
\label{eq:S1_PAM_quadratic}
\end{equation}
Again, $\mathcal P_{AM}$ is the positive root of Eq.~\eqref{eq:S1_PAM_quadratic}.

Thus, the equations and thresholds used in the main text are recovered as a particular specialization of the broader modal framework developed in this Supplementary Note.
\section*{Supplementary Note 2: Three-mode system}

In this Supplementary Note we extend the general two-mode reduction of Supplementary Note~1 to the case of three interacting modes: the ground mode and the two degenerate vortex modes with azimuthal quantum numbers $m=\pm1$. This three-mode model is used in the main text to compare the full numerical simulations with the reduced modal dynamics during the initial and intermediate stages of condensate formation.

As in Supplementary Note~1, we work in a generalized basis $\varphi_j$ normalized in the same way as the conservative-trap basis used in the main text. The difference between the two descriptions lies in the choice of basis functions: the generalized modes $\varphi_j$ may include the pump-induced linear blueshift, whereas the modes $\phi_j$ used in the main text are eigenmodes of the conservative trap.

\subsection*{A. Effective condensate equation}

We start from the effective driven-dissipative condensate equation derived in Supplementary Note~1,
\begin{equation}
i\partial_t\psi
=
\left[
-\frac{\hbar \nabla^2}{2m_{LP}}
+V_{\mathrm{trap}}(r)
+\frac{2\tilde g\,P(\mathbf r)}{\gamma_R}
\right]\psi
+\bigl(g_{\mathrm{eff}}(\mathbf r)-i\chi(\mathbf r)\bigr)|\psi|^2\psi
+i\Gamma_0(\mathbf r)\psi
+\frac{i\lambda}{2}\psi\left(\psi^*\nabla^2\psi-\psi\nabla^2\psi^*\right),
\label{eq:S2_effective_GPE}
\end{equation}
where
\begin{align}
g_{\mathrm{eff}}(\mathbf r)
&=
g-\frac{2\tilde g\,R}{\gamma_R^2}P(\mathbf r),
\\
\chi(\mathbf r)
&=
\frac{R^2}{2\gamma_R^2}P(\mathbf r),
\\
\Gamma_0(\mathbf r)
&=
\frac{1}{2}\left(\frac{R\,P(\mathbf r)}{\gamma_R}-\gamma\right).
\end{align}

We assume radial symmetry of both the trap and the pump profile. The relevant low-lying modes are then the fundamental mode and the two degenerate vortex modes with angular momenta $m=\pm1$.

\subsection*{B. Three-mode expansion}

We expand the condensate field in the three modes
\begin{equation}
\psi(r,\theta,t)
=
B_0(t)\,\varphi_0(r,\theta)\,e^{-i\omega_0 t}
+
B_+(t)\,\varphi_+(r,\theta)\,e^{-i\omega_v t}
+
B_-(t)\,\varphi_-(r,\theta)\,e^{-i\omega_v t},
\label{eq:S2_ansatz}
\end{equation}
where
\begin{equation}
\varphi_0(r,\theta)=u_0(r),
\qquad
\varphi_+(r,\theta)=u_v(r)e^{+i\theta},
\qquad
\varphi_-(r,\theta)=u_v(r)e^{-i\theta},
\label{eq:S2_modes}
\end{equation}
and $\omega_v$ is the common linear eigenfrequency of the degenerate vortex pair.

The generalized modes are normalized as
\begin{equation}
\int_{\mathbb R^2}\varphi_i^*(\mathbf r)\varphi_j(\mathbf r)\,d^2r=\delta_{ij},
\qquad
i,j\in\{0,+,-\}.
\label{eq:S2_norm}
\end{equation}
For the radial functions introduced in Eq.~\eqref{eq:S2_modes}, this normalization is equivalent to
\begin{equation}
2\pi\int_0^\infty u_j^2(r)\,r\,dr=1,
\qquad
j\in\{0,+,-\}.
\end{equation}

The modal amplitudes are defined by projection,
\begin{equation}
B_j(t)e^{-i\omega_j t}
=
\int_{\mathbb R^2}\varphi_j^*(\mathbf r)\,\psi(\mathbf r,t)\,d^2r,
\qquad
j\in\{0,+,-\},
\label{eq:S2_projection}
\end{equation}
with $\omega_+=\omega_-=\omega_v$.

Because the pump and trap are radially symmetric, the two vortex modes have identical radial profiles and identical linear gains. In particular,
\begin{equation}
\bar\Gamma_+=\bar\Gamma_-=\bar\Gamma_v,
\qquad
|\varphi_+|^2=|\varphi_-|^2\equiv |\varphi_v|^2.
\label{eq:S2_symmetry_linear}
\end{equation}

\subsection*{C. Nonlinear coefficients}

The net linear gains of the modes are determined by their overlap with the spatial gain profile:
\begin{equation}
\bar\Gamma_j
=
\int_{\mathbb R^2}\Gamma_0(r)\,|\varphi_j(\mathbf r)|^2\,d^2r,
\qquad
j\in\{0,+,-\}.
\label{eq:S2_Gammabar}
\end{equation}

The conservative nonlinear coefficients are
\begin{equation}
\bar u_{ij}
=
\int_{\mathbb R^2}g_{\mathrm{eff}}(r)\,|\varphi_i|^2|\varphi_j|^2\,d^2r,
\qquad
i,j\in\{0,+,-\},
\label{eq:S2_ubar}
\end{equation}
and the dissipative saturation coefficients are
\begin{equation}
\bar\sigma_{ij}
=
\int_{\mathbb R^2}\chi(r)\,|\varphi_i|^2|\varphi_j|^2\,d^2r,
\qquad
i,j\in\{0,+,-\}.
\label{eq:S2_sigmabar}
\end{equation}

The relaxation-induced transfer between the ground mode and each of the vortex modes is described by
\begin{equation}
\bar\rho
\equiv
\lambda\,\frac{m_{LP}}{\hbar}\,(\omega_v-\omega_0)\,
\int_{\mathbb R^2}|\varphi_0|^2|\varphi_v|^2\,d^2r.
\label{eq:S2_rho}
\end{equation}
By radial symmetry one has
\begin{equation}
\bar\rho_{+0}=\bar\rho_{-0}\equiv \bar\rho.
\end{equation}
No relaxation-induced transfer appears between the two vortex modes themselves, because they are degenerate:
\begin{equation}
\omega_+-\omega_-=0.
\end{equation}

For later convenience, we note the symmetry relations
\begin{align}
\bar u_{0+}&=\bar u_{0-}\equiv \bar u_{0v},
&
\bar u_{++}&=\bar u_{--}=\bar u_{+-}=\bar u_{-+}\equiv \bar u_{vv},
\label{eq:S2_u_sym}
\\
\bar\sigma_{0+}&=\bar\sigma_{0-}\equiv \bar\sigma_{0v},
&
\bar\sigma_{++}&=\bar\sigma_{--}=\bar\sigma_{+-}=\bar\sigma_{-+}\equiv \bar\sigma_{vv}.
\label{eq:S2_sigma_sym}
\end{align}

\subsection*{D. Coupled equations for the mode amplitudes}

Substituting the ansatz \eqref{eq:S2_ansatz} into Eq.~\eqref{eq:S2_effective_GPE}, projecting onto $\varphi_0 e^{i\omega_0 t}$ and $\varphi_{\pm} e^{i\omega_v t}$, and retaining only resonant terms within the rotating-wave approximation, we obtain the coupled equations for the modal amplitudes.

The ground-mode amplitude obeys
\begin{align}
i\,\dot B_0
&=
i\,\bar\Gamma_0 B_0
+
(\bar u_{00}-i\bar\sigma_{00})|B_0|^2 B_0
+
2(\bar u_{0v}-i\bar\sigma_{0v})|B_+|^2 B_0
\nonumber\\
&\hspace{0.8cm}
+
2(\bar u_{0v}-i\bar\sigma_{0v})|B_-|^2 B_0
+
i\bar\rho\,|B_+|^2 B_0
+
i\bar\rho\,|B_-|^2 B_0 .
\label{eq:S2_B0_amp}
\end{align}
The amplitudes of the two vortex modes satisfy
\begin{align}
i\,\dot B_+
&=
i\,\bar\Gamma_v B_+
+
(\bar u_{vv}-i\bar\sigma_{vv})|B_+|^2 B_+
+
2(\bar u_{0v}-i\bar\sigma_{0v})|B_0|^2 B_+
\nonumber\\
&\hspace{0.8cm}
+
2(\bar u_{vv}-i\bar\sigma_{vv})|B_-|^2 B_+
-
i\bar\rho\,|B_0|^2 B_+ ,
\label{eq:S2_Bplus_amp}
\\
i\,\dot B_-
&=
i\,\bar\Gamma_v B_-
+
(\bar u_{vv}-i\bar\sigma_{vv})|B_-|^2 B_-
+
2(\bar u_{0v}-i\bar\sigma_{0v})|B_0|^2 B_-
\nonumber\\
&\hspace{0.8cm}
+
2(\bar u_{vv}-i\bar\sigma_{vv})|B_+|^2 B_-
-
i\bar\rho\,|B_0|^2 B_- .
\label{eq:S2_Bminus_amp}
\end{align}

Dividing by $i$ and factoring out the corresponding amplitude, we rewrite the system as
\begin{align}
\dot B_0
&=
\Big[
\bar\Gamma_0
-\bar\sigma_{00}|B_0|^2
-(2\bar\sigma_{0v}-\bar\rho)|B_+|^2
-(2\bar\sigma_{0v}-\bar\rho)|B_-|^2
\Big]B_0
\nonumber\\
&\hspace{0.8cm}
-i\Big[
\bar u_{00}|B_0|^2
+2\bar u_{0v}|B_+|^2
+2\bar u_{0v}|B_-|^2
\Big]B_0,
\label{eq:S2_B0_normal}
\\[4pt]
\dot B_+
&=
\Big[
\bar\Gamma_v
-\bar\sigma_{vv}|B_+|^2
-(2\bar\sigma_{0v}+\bar\rho)|B_0|^2
-2\bar\sigma_{vv}|B_-|^2
\Big]B_+
\nonumber\\
&\hspace{0.8cm}
-i\Big[
\bar u_{vv}|B_+|^2
+2\bar u_{0v}|B_0|^2
+2\bar u_{vv}|B_-|^2
\Big]B_+,
\label{eq:S2_Bplus_normal}
\\[4pt]
\dot B_-
&=
\Big[
\bar\Gamma_v
-\bar\sigma_{vv}|B_-|^2
-(2\bar\sigma_{0v}+\bar\rho)|B_0|^2
-2\bar\sigma_{vv}|B_+|^2
\Big]B_-
\nonumber\\
&\hspace{0.8cm}
-i\Big[
\bar u_{vv}|B_-|^2
+2\bar u_{0v}|B_0|^2
+2\bar u_{vv}|B_+|^2
\Big]B_-.
\label{eq:S2_Bminus_normal}
\end{align}

As in the two-mode case, the conservative coefficients $\bar u_{ij}$ affect only the nonlinear phase evolution and do not enter directly into the intensity equations.

\subsection*{E. Intensity equations}

Introducing the modal intensities
\begin{equation}
\mathcal I_0=|B_0|^2,
\qquad
\mathcal I_+=|B_+|^2,
\qquad
\mathcal I_-=|B_-|^2,
\label{eq:S2_intensities}
\end{equation}
we obtain
\begin{align}
\dot{\mathcal I}_0
&=
2\mathcal I_0\Big[
\bar\Gamma_0
-\bar\sigma_{00}\mathcal I_0
-(2\bar\sigma_{0v}-\bar\rho)\mathcal I_+
-(2\bar\sigma_{0v}-\bar\rho)\mathcal I_-
\Big],
\label{eq:S2_I0}
\\[4pt]
\dot{\mathcal I}_+
&=
2\mathcal I_+\Big[
\bar\Gamma_v
-\bar\sigma_{vv}\mathcal I_+
-(2\bar\sigma_{0v}+\bar\rho)\mathcal I_0
-2\bar\sigma_{vv}\mathcal I_-
\Big],
\label{eq:S2_Iplus}
\\[4pt]
\dot{\mathcal I}_-
&=
2\mathcal I_-\Big[
\bar\Gamma_v
-\bar\sigma_{vv}\mathcal I_-
-(2\bar\sigma_{0v}+\bar\rho)\mathcal I_0
-2\bar\sigma_{vv}\mathcal I_+
\Big].
\label{eq:S2_Iminus}
\end{align}

These equations show explicitly that the two vortex components compete through mutual cross-saturation with the same coefficient as their self-saturation. As a result, once the linear growth stage ends, one of the two vortex modes typically suppresses the other and only one of them survives. This is why the late-time dynamics may be described by the reduced two-mode model involving only the ground state and a single vortex mode.

\subsection*{F. Relation to the notation used in the main text}

The main text uses orthonormal conservative-trap modes $\phi_j$ and amplitudes $A_j$. In the present Supplementary Note, the generalized modes $\varphi_j$ are normalized in the same way:
\begin{equation}
\int_{\mathbb R^2}\phi_i^*(\mathbf r)\phi_j(\mathbf r)\,d^2r
=
\int_{\mathbb R^2}\varphi_i^*(\mathbf r)\varphi_j(\mathbf r)\,d^2r
=
\delta_{ij}.
\end{equation}
However, the two bases correspond to different linear problems. The modes $\phi_j$ are eigenmodes of the conservative trap, whereas the generalized modes $\varphi_j$ may incorporate the pump-induced linear blueshift. Therefore, in general, the amplitudes $A_j$ and $B_j$ are different projection coefficients.

In the weak-blueshift regime, after factorizing the pump as
\begin{equation}
P(\mathbf r)=\mathcal P\,\eta(r),
\qquad
\max_r\eta(r)=1,
\end{equation}
the generalized basis reduces to the conservative-trap basis, and the equations above reduce to the three-mode ODE system used in the main text to compare the reduced modal dynamics with the full numerical simulations. In this specialization, the effective linear gains take the form
\begin{equation}
\Gamma_j=-\tilde\Gamma_j+\epsilon_j\mathcal P,
\end{equation}
while the saturation coefficients become proportional to $\mathcal P$, in complete analogy with the two-mode reduction discussed in Supplementary Note~1.

Thus, the three-mode equations derived here provide the general modal framework underlying the reduced ODE description used in the main text for the dynamics of the modes $m=0$ and $m=\pm1$.

\section*{Supplementary Note 3: Three-mode dynamics within the perturbative theory}

To complement the direct numerical simulations shown in the main text, we also solved the reduced three-mode system obtained within the perturbative approach. In this description, the condensate is represented by the amplitudes of the ground mode and the two degenerate vortex modes with azimuthal indices $m=\pm1$. The corresponding intensity equations are given in Supplementary Note~2.

\begin{figure}[t!]
\includegraphics[width=0.7\linewidth]{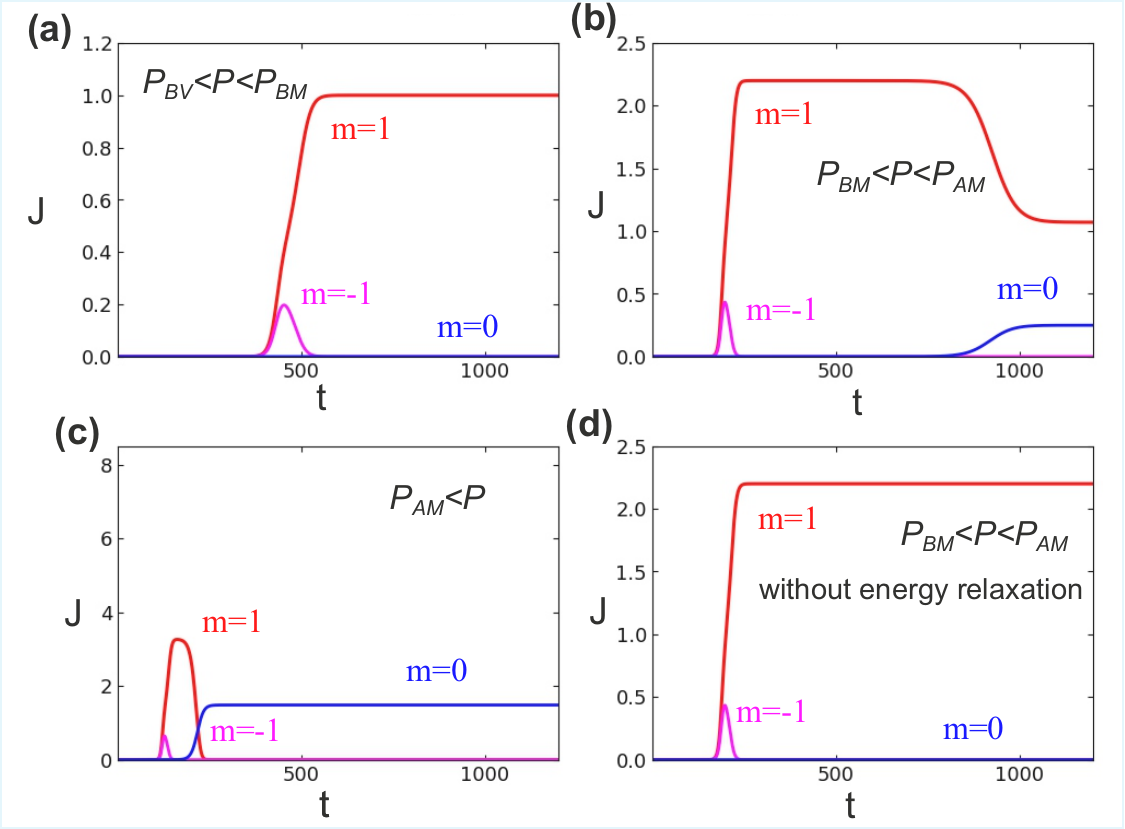}
\caption{Dynamics of the modal intensities within the reduced three-mode perturbative model. The panels correspond to the same pump values as in Fig.~2 of the main text. Panels (a)--(c) show the case with energy relaxation for $ \mathcal P=1.106$, $ \mathcal P=1.172$, and $ \mathcal P=1.238$, respectively, while panel (d) shows the same pump as in panel (b), but without energy relaxation. The blue curves correspond to the ground-state mode with $m=0$, the red curves to the vortex mode with $m=1$, and the magenta curves to the vortex mode with $m=-1$. The reduced model reproduces the same qualitative evolution as the direct numerical simulations: for $\mathcal P_{BV}< \mathcal P< \mathcal P_{BM}$ a vortex state is selected, for $\mathcal P_{BM}< \mathcal P< \mathcal P_{AM}$ a mixed regime develops, and for $\mathcal P> \mathcal P_{AM}$ the ground-state mode becomes dominant, whereas without energy relaxation the condensate remains vortex dominated.}
\label{fig:perturbative_dynamics}
\end{figure}

Figure~S\ref{fig:perturbative_dynamics} shows the temporal evolution of the modal intensities for the same representative pump values as in the main text. Small asymmetric initial conditions were used in order to mimic weak symmetry breaking and allow the system to select one of the two vortex components. The pump thresholds $\mathcal P_{BM}$ and $\mathcal P_{AM}$ were evaluated from the perturbative theory itself.

The reduced model reproduces the same qualitative scenario as the full simulations. For $\mathcal P_{BV}< \mathcal P< \mathcal P_{BM}$, one of the vortex modes suppresses its counterpart and the system evolves to a vortex state. In the interval $\mathcal P_{BM}< \mathcal P< \mathcal P_{AM}$, the initially selected vortex state subsequently induces the growth of the ground-state component, leading to a mixed regime. For $ \mathcal P> \mathcal P_{AM}$, the ground-state mode ultimately dominates the dynamics. By contrast, when the relaxation-induced transfer is switched off, the ground-state component does not develop and the condensate remains vortex-dominated. Thus, the perturbative theory captures not only the stationary-state selection but also the transient route by which the system approaches the asymptotic regime.

\section*{Supplementary Note 4: Dimensionless equation used in the numerics and conversion to physical units}

The numerical simulations shown in the main text were performed using the dimensionless equation
\begin{equation}
\partial_{\tilde t}\psi
=
\frac{i}{2}\tilde\nabla^2\psi
-\gamma_d\psi
+\bigl[W(\tilde{\mathbf r})-i\bigl(V(\tilde{\mathbf r})+\beta W(\tilde{\mathbf r})\bigr)\bigr]\psi
-i|\psi|^2\psi
-\mu W(\tilde{\mathbf r})(1-i\beta)|\psi|^2\psi
+\frac{\lambda}{2}\psi\left(\psi^*\tilde\nabla^2\psi-\psi\tilde\nabla^2\psi^*\right).
\label{eq:S4_dimless_eq}
\end{equation}
Here tildes denote dimensionless variables, $\gamma_d$ is the dimensionless linear decay coefficient, and $\lambda$ is the dimensionless pure energy-relaxation coefficient.

This equation is the dimensionless form of the effective model used in the main text. The correspondence between the parameters in Eq.~\eqref{eq:S4_dimless_eq} and the notation of the main text is
\begin{equation}
V(\tilde r)=\tau\,V_{\mathrm{trap}}(x_0\tilde r),
\qquad
W(\tilde r)=\tau\,\frac{R\mathcal P}{2\gamma_R}\,\eta(x_0\tilde r),
\label{eq:S4_VW_def}
\end{equation}
\begin{equation}
\gamma_d=\frac{\tau\gamma}{2},
\qquad
\mu=\frac{R\rho_0}{\gamma_R},
\qquad
\beta=\frac{4\tilde g}{R},
\qquad
\lambda=\lambda_{\mathrm{phys}}\frac{\rho_0\tau}{x_0^2},
\label{eq:S4_parameter_map}
\end{equation}
where $\gamma$ and $\lambda_{\mathrm{phys}}$ are the dimensional coefficients entering the main-text equation.

To convert the dimensionless parameters to physical units, we introduce the characteristic density $\rho_0$, the interaction energy scale
\begin{equation}
E_0=g_E\rho_0,
\end{equation}
the time scale
\begin{equation}
\tau=\frac{\hbar}{E_0},
\end{equation}
and the length scale
\begin{equation}
x_0=\sqrt{\frac{\hbar^2}{m_{LP}E_0}}.
\end{equation}
Here $g_E$ is the polariton-polariton interaction constant in energy units. The dimensional and dimensionless variables are related by
\begin{equation}
\Psi(\mathbf r,t)=\sqrt{\rho_0}\,\psi(\tilde{\mathbf r},\tilde t),
\qquad
\mathbf r=x_0\tilde{\mathbf r},
\qquad
t=\tau \tilde t.
\label{eq:S4_scaling}
\end{equation}

We use the following physical input parameters:
\begin{equation}
g_E=6\times10^{-3}\ \mathrm{meV\,\mu m^2},
\qquad
\rho_0=50\ \mathrm{\mu m^{-2}},
\qquad
m_{LP}=5\times10^{-5}m_e,
\qquad
\hbar=0.658\ \mathrm{meV\,ps}.
\end{equation}
They give
\begin{align}
E_0&=g_E\rho_0=0.3\ \mathrm{meV},
\\
\tau&=\frac{\hbar}{E_0}\approx 2.194\ \mathrm{ps},
\\
x_0&=\sqrt{\frac{\hbar^2}{m_{LP}E_0}}\approx 2.254\ \mathrm{\mu m},
\\
\sqrt{\rho_0}&=\sqrt{50}\approx 7.071\ \mathrm{\mu m^{-1}}.
\end{align}

The dimensionless parameters used in the simulations are
\begin{equation}
\begin{aligned}
\gamma_d &= \frac12, &
\mu &= 0.05, &
\beta &= 2.5, &
\lambda &= 0.0025, &
V_0 &= -15, & \\
R_0 &= 1.25, & 
h_0 &= 0.75, &
R_1 &= 0.8, &
h_1 &= 0.3536, &
W_0 &\ \text{varied}.
\end{aligned}
\end{equation}
For the simulations without energy relaxation, one sets $\lambda=0$.

The conservative potential is taken in the form
\begin{equation}
V(\tilde r)=
V_0\frac{1}{1+\exp\!\left(\frac{\tilde r-R_0}{h_0}\right)},
\qquad
V_0<0,
\label{eq:S4_potential}
\end{equation}
while the gain profile is
\begin{equation}
W(\tilde r)=
W_0\left[
\exp\!\left(-\frac{(\tilde r-R_1)^2}{h_1^2}\right)
+
\exp\!\left(-\frac{(\tilde r+R_1)^2}{h_1^2}\right)
\right].
\label{eq:S4_gain}
\end{equation}
The parameter $W_0$ is proportional to the physical pump amplitude $\mathcal P$ [see Eq.~\eqref{eq:S4_VW_def}], so varying $W_0$ in the simulations is equivalent to varying $\mathcal P$ in the notation of the main text. A representative value in the pump range considered in the main text is $W_0\simeq 1.16$.

The corresponding dimensional parameters are
\begin{align}
\gamma_{\rm phys}
&=
\frac{2\gamma_d}{\tau}
=
\frac{2\times 0.5}{2.194}
\approx 0.456\ \mathrm{ps^{-1}},
\\[1mm]
V_{0,\rm phys}^{(\mathrm{energy})}
&=
E_0V_0
=
0.3\times (-15)
=
-4.5\ \mathrm{meV},
\\
V_{0,\rm phys}^{(\mathrm{freq})}
&=
\frac{V_0}{\tau}
=
\frac{-15}{2.194}
\approx -6.84\ \mathrm{ps^{-1}},
\\[1mm]
W_{0,\rm phys}
&=
\frac{W_0}{\tau}.
\end{align}
For $W_0\simeq 1.16$, this gives
\begin{equation}
W_{0,\rm phys}\approx \frac{1.16}{2.194}\approx 0.529\ \mathrm{ps^{-1}}.
\end{equation}
The geometric parameters become
\begin{align}
R_{0,\rm phys}
&=
x_0R_0
=
2.254\times1.25
\approx 2.82\ \mathrm{\mu m},
\\
h_{0,\rm phys}
&=
x_0h_0
=
2.254\times0.75
\approx 1.69\ \mathrm{\mu m},
\\
R_{1,\rm phys}
&=
x_0R_1
=
2.254\times0.8
\approx 1.80\ \mathrm{\mu m},
\\
h_{1,\rm phys}
&=
x_0 h_1
=
2.254\times 0.3536
\approx 0.797\ \mathrm{\mu m}.
\end{align}
Finally, the dimensional pure energy-relaxation coefficient is
\begin{equation}
\lambda_{\rm phys}
=
\lambda\,\frac{x_0^2}{\rho_0\tau}
=
0.0025\times\frac{(2.254)^2}{50\times 2.194}
\approx 1.16\times 10^{-4}\ \mathrm{\mu m^4/ps}.
\label{eq:S4_lambda_phys}
\end{equation}